\documentclass[aps,prd,10pt,tightenlines,amsmath,unsortedaddress,twocolumn,amssymb,10pt]{revtex4}

\usepackage{float}
\usepackage{graphicx}% Include figure*ure files
\usepackage{dcolumn}% Align table columns on decimal point
\usepackage{bm}% bold math
\usepackage{bbold}
\usepackage{enumerate}
\usepackage[utf8]{inputenc}
\usepackage[english]{babel}
\usepackage{amsthm}
\usepackage{xcolor}

\begin{document}

\title{ALP induced polarization effects on photons from blazars}

\author{Giorgio Galanti}
\email{gam.galanti@gmail.com}
\affiliation{INAF, Istituto di Astrofisica Spaziale e Fisica Cosmica di Milano, Via Alfonso Corti 12, I -- 20133 Milano, Italy}

\author{Marco Roncadelli}
\email{marcoroncadelli@gmail.com}
\affiliation{INFN, Sezione di Pavia, Via Agostino Bassi 6, I -- 27100 Pavia, Italy}
\affiliation{INAF, Osservatorio Astronomico di Brera, Via Emilio Bianchi 46, I -- 23807 Merate, Italy}

\author{Fabrizio Tavecchio}
\email{fabrizio.tavecchio@inaf.it}
\affiliation{INAF, Osservatorio Astronomico di Brera, Via Emilio Bianchi 46, I -- 23807 Merate, Italy}

\date{\today}

\begin{abstract}

Axion-like particles (ALPs), which are very light neutral spin zero elusive particles primarily interacting with two photons and predicted by superstring and superbrane theories, have come to help by solving two distinct problems about blazars (a type of active galactic nuclei), thus providing two hints at the existence of ALPs themselves. In the presence of an external magnetic field, ALPs produce: (i) photon-ALP oscillations, (ii) the change of the polarization state of photons. The former effect has many consequences in the astrophysical contest, such as the modification of the transparency of the Universe and the alteration of the astrophysical spectra. We address here the latter effect by analyzing how the photon degree of linear polarization and the polarization angle get modified by photon-ALP interaction in the case where photons are generated at the jet base of some BL~Lacs (a blazar class): OJ~287, BL~Lacertae, Markarian~501 and 1ES~0229+200, by considering both a leptonic and hadronic emission mechanism. We show that OJ~287 and BL~Lacertae are good observational targets for ALP studies both in the X-ray band with IXPE (already operative) and with the proposed eXTP, XL-Calibur, NGXP and XPP missions and in the high-energy range with the COSI, e-ASTROGAM and AMEGO missions, while 1ES~0229+200 represents a strong candidate in the X-ray band only. Since these blazars show a very high final photon degree of linear polarization, which cannot be explained by conventional physics, such a possible detection would represent an additional hint at the ALP existence. Instead, Markarian~501 does not appear as a good target for these studies. We conclude that all these observatories can give us additional fundamental information about ALP physics.

\end{abstract}



\maketitle

%%%%%%%%%%%%%%%%%%%%%%%%%%%%%%%%%%%%%%%%%%%%
%% MAINMATTER
%%%%%%%%%%%%%%%%%%%%%%%%%%%%%%%%%%%%%%%%%%%%

%\newpage

\section{Introduction}
Several extensions of the standard model of particle physics, such as the superstring and superbrane theories~\cite{string1,string2,string3,string4,string5,axiverse,abk2010,cicoli2012,cisterna1,cisterna2}, invariably predict the existence of axion-like particles (ALPs, see e.g.~\cite{alp1,alp2}). ALPs are very light neutral pseudoscalar bosons and are nowadays considered among the best candidates to constitute the dark matter~\cite{preskill,abbott,dine,arias2012,jaekel}. ALPs are a generalization of the axion, namely the pseudo-Goldsone boson deriving from the breakdown of the global Peccei-Quinn symmetry $\rm U(1)_{PQ}$ proposed as a solution to the strong CP problem~\cite{axionrev1,axionrev2,axionrev3,axionrev4}. However, ALPs differ from the original axion into two main aspects: (i) the ALP mass $m_a$ and the ALP-to-two-photon coupling $g_{a\gamma\gamma}$ are uncorrelated quantities while they are linked in the case of the axion, (ii) ALPs primarily interact with two photons, while other interactions with fermions and gluons -- which are instead fundamental in the case of the axion -- are subdominant. When an external magnetic field is present, two effects are produced: (i) photon-ALP oscillations~\cite{mpz,raffeltstodolsky}, (ii) the change of the polarization state of photons~\cite{mpz,raffeltstodolsky}.

ALP detection is made difficult by their very faint interaction with photons over distances attainable in laboratory experiments. Instead, since the astrophysical environment is not affected by this limitation, it represents the best opportunity to study ALP physics especially in the high-energy (HE) and very-high-energy (VHE) bands, where the former above-mentioned ALP effect, i.e.  photon-ALP oscillation, produces many consequences (see e.g.~\cite{irastorzaredondo,gRew,grRew}). At the same time, VHE astrophysics presents several challenging issues, for the solution of which the inclusion of ALPs have repeatedly been invoked. In particular, two {\it hints} at the ALP existence arise from blazars, which are a class of active galactic nuclei (AGN). Specifically, ALPs naturally explain why flat spectrum radio quasars (a blazar subclass) are observed up to $400 \, \rm GeV$, in spite of the fact that conventional physics prevents any emission above $30 \, \rm GeV$~\cite{trgb2012} -- which represents a first hint at the existence of an ALP. In addition, ALPs solve the anomalous redshift dependence of the {\it emitted} spectra of BL~Lacs (a blazar subclass)~\cite{grdb} -- which represents a second hint. Photon-ALP interaction is also responsible for the alteration of the photon transparency (see e.g.~\cite{drm2007,dgr2011}) and give rise to an oscillatory behavior in observed spectra~\cite{dmr2008,wb2012,gr2013,grExt,gtre2019,gtl2020}. The detection of the gamma-ray burst GRB 221009A at $18 \, \rm TeV$ by LHAASO~\cite{LHAASO} and even at $251 \, \rm TeV$ by Carpet-2~\cite{carpet} suggests a strong indication for the existence of an ALP with the properties used in the previous two hints~\cite{grtGRB}.

Also the latter above-mentioned ALP effect, i.e. the change of the polarization state of photons, has sizable and detectable consequences. In particular, the polarization of photons from gamma-ray bursts has been studied in~\cite{bassan}, while other studies 
cover different topics concerning several astrophysical sources~\cite{ALPpol1,ALPpol2,ALPpol3,ALPpol4,ALPpol5,day}. In addition, the photon-ALP interaction can be employed to measure {\it emitted} photon polarization~\cite{galantiTheo}. Very recently, ALP-induced polarization effects have been analyzed when photons are generated in the central region of galaxy clusters or in the blazar jet~\cite{galantiPol}. In particular, the Perseus and Coma clusters have been identified as solid targets for studies concerning ALP effects on photon polarization~\cite{grtClu}. Because of the new interest on photon polarization, as shown by the launch or proposal of new observatories in the X-ray band like IXPE~\cite{ixpe}, eXTP~\cite{extp}, XL-Calibur~\cite{xcalibur}, NGXP~\cite{ngxp} and XPP~\cite{xpp} and in the HE band such as COSI~\cite{cosi}, e-ASTROGAM~\cite{eastrogam1,eastrogam2} and AMEGO~\cite{amego}, in this paper we consider peculiar BL~Lacs in order to understand which ones are the best observational targets for ALP studies.

In particular, by closely following the analysis developed in~\cite{galantiPol}, we investigate the photon-ALP beam propagation for photons emitted at the jet base of several BL~Lacs: OJ~287, BL~Lacertae, Markarian~501 and 1ES~0229+200 and we compute -- for each of them -- the photon survival probability $P_{\gamma \to \gamma}$ and the corresponding final photon degree of linear polarization $\Pi_L$ and polarization angle $\chi$. By using the state-of-the-art knowledge concerning the crossed regions (blazar jet, host galaxy, extragalactic space, Milky Way) and physically consistent bounds on the photon-ALP system parameters, we conclude that ALPs induce sizable effects on the final photon polarization which is observable by present and planned missions~\cite{ixpe,extp,xcalibur,ngxp,xpp,cosi,eastrogam1,eastrogam2,amego}. In particular, OJ~287 and BL~Lacertae appear as good observational targets both in the X-ray and in the HE band, 1ES~0229+200 in the X-ray range only, while Markarian~501 does not represent a good candidate for studies of ALP effects on photon polarization.

The paper is organized as follows. In Sec. II we introduce ALPs and polarization effects, in Sec. III we cursorily describe the properties of the media crossed by the photon-ALP beam, in Sec. IV we present our results, while in Sec. V we draw our conclusions.

\section{Axion-like particles and polarization effects}

ALPs are very light neutral pseudo-scalar bosons, whose primary interaction with two photons is governed by the Lagrangian
\begin{eqnarray}
&\displaystyle {\cal L}_{\rm ALP} =  \frac{1}{2} \, \partial^{\mu} a \, \partial_{\mu} a - \frac{1}{2} \, m_a^2 \, a^2 - \, \frac{1}{4 } g_{a\gamma\gamma} \, F_{\mu\nu} \tilde{F}^{\mu\nu} a \nonumber \\
&\displaystyle = \frac{1}{2} \, \partial^{\mu} a \, \partial_{\mu} a - \frac{1}{2} \, m_a^2 \, a^2 + g_{a\gamma\gamma} \, {\bf E} \cdot {\bf B}~a~,
\label{lagr}
\end{eqnarray}
where $a$ is the ALP field, while ${\bf E}$ and ${\bf B}$ express the electric and magnetic parts of the electromagnetic tensor $F_{\mu\nu}$, whose dual is represented by $\tilde{F}^{\mu\nu}$. In many VHE studies in addition to Eq.~(\ref{lagr}) also the Heisenberg-Euler-Weisskopf (HEW) effective Lagrangian
\begin{equation}
\label{HEW}
{\cal L}_{\rm HEW} = \frac{2 \alpha^2}{45 m_e^4} \, \left[ \left({\bf E}^2 - {\bf B}^2 \right)^2 + 7 \left({\bf E} \cdot {\bf B} \right)^2 \right]~,
\end{equation}
must be considered, where $\alpha$ is the fine-structure constant, while $m_e$ is the electron mass~\cite{hew1, hew2, hew3}. Eq.~(\ref{HEW}) describes the photon one-loop vacuum polarization. This effect can be relevant in the case of hadronic emission models for the very high central jet magnetic field (see also~\cite{grSM,grExt} and Sec. III). 

Many bounds on ALP parameters ($m_a, g_{a\gamma\gamma}$) exist in the literature~\cite{cast,straniero,fermi2016,payez2015,berg,conlonLim,meyer2020,limFabian,limJulia,limKripp,limRey2,mwd}, however, the most reliable one is that of CAST,  arising from no-detection of ALPs from the Sun and reads $g_{a \gamma \gamma} < 0.66 \times 10^{- 10} \, {\rm GeV}^{- 1}$ for $m_a < 0.02 \, {\rm eV}$ at the $2 \sigma$ level~\cite{cast}.

We consider a photon-ALP beam of energy $E$, propagating in several magnetized media (blazar jet, host galaxy, extragalactic space, Milky Way) up to its arrival at the Earth, where photons can be detected. We do not report here the full details of the calculation of the photon survival probability $P_{\gamma \to \gamma}$ and of the corresponding photon degree of linear polarization $\Pi_L$ and polarization angle $\chi$, and we refer the reader to~\cite{galantiPol,grtClu}, where a full description of the theoretical derivation and calculation scheme is reported. We just want to recall that we take all the magnetization and absorption properties of the crossed media into account by using the state-of-the-art knowledge (see also~\cite{galantiPol,grtClu} and Sec. III for more details).

We conclude this section by mentioning that a strict relationship between the {\it emitted} photon degree of linear polarization $\Pi_{L,0}$ and the photon survival probability $P_{\gamma \to \gamma}$ in the absence of photon absorption has been demonstrated in~\cite{galantiTheo}. In particular, we have $P_{\gamma \to \gamma} \geq (1-\Pi_{L,0})/2$. Among other important implications, the previous inequality derived in~\cite{galantiTheo} represents a check for the correctness of our results. We have checked that $P_{\gamma \to \gamma}$ satisfies the latter relation in all the following figures.

\section{Photon-ALP beam propagation} 

In the following, we sketchily illustrate the main properties of the astrophysical media crossed by the photon-ALP beam (blazar jet, host galaxy, extragalactic space, Milky Way), by underlining those that are more important for the photon-ALP system. For a complete description, we address the reader also to~\cite{galantiPol,grtClu} and to the specific below-cited publications dealing with the particular topic. Concerning the photon-ALP system parameters, we assume the benchmark values $g_{a\gamma\gamma}= 0.5 \times 10^{-11} \, \rm GeV^{-1}$ and the two ALP masses: (i) $m_a \lesssim 10^{-14} \, \rm eV$ and (ii) $m_a = 10^{-10} \, \rm eV$. This choice allow us to accomplish the firmest ALP bound by CAST~\cite{cast}.

\subsection{Blazar}

Blazars are a class of AGN, namely extragalactic supermassive black holes (SMBHs), which efficiently accrete matter from the surrounding. In such a situation, two collimated relativistic jets are emitted in opposite directions. When, just by chance, one of them turns out to be in the direction of the Earth, the AGN are called a blazar. In the following, we deal with a subclass of blazars called BL~Lac objects (BL~Lacs). BL~Lacs are less powerful than the objects of the other class called flat spectrum radio quasars (FSRQs), and at variance with FSRQs lack strong optical emission lines and VHE absorption regions.

Photons are generated at the BL~Lac emission region, which lies at a distance of about $y_{\rm em}= (10^{16} - 10^{17}) \, {\rm cm}$ from the central SMBH. Then, photons can oscillate into ALPs in the magnetic field of the jet ${\bf B}^{\rm jet}$ until it ends at a distance of about $1 \, \rm kpc$. Here, the photon-ALP beam enters the host galaxy. Since we are far enough from the center, the toroidal part of ${\bf B}^{\rm jet}$ -- transverse to the jet axis~\cite{bbr1984,ghisellini2009,pudritz2011} -- is dominant and its profile reads 
\begin{equation}
\label{Bjet}
B^{\rm jet} ( y ) = B^{\rm jet}_0 \left(\frac{y_{{\rm em}}}{y}\right)~,
\end{equation}
where $B^{\rm jet}_0$ represents the jet magnetic field strength at $y_{\rm em}$. The electron number density $n_e^{\rm jet}$ profile is expressed by 
\begin{equation}
\label{njet}
n^{\rm jet}_e ( y ) = n^{\rm jet}_{e,0} \left(\frac{y_{{\rm em}}}{y}\right)^2~,
\end{equation}
for the conical shape of the jet. In Eq.~(\ref{njet}) $n^{\rm jet}_{e,0}$ accounts for the jet electron number density at $y_{\rm em}$. Synchrotron Self Compton (SSC) diagnostics applied to blazar spectra suggests us to take $n^{\rm jet}_{e,0}=5 \times 10^4 \, \rm cm^{-3}$~\cite{tavecchio2010}.

In the following, we consider both a leptonic and a hadronic emission mechanism. In either case photons in the optical up to the X-ray band are generated by electron-synchrotron emission, while at higher energies the leptonic model generates photons via inverse Compton scattering~\cite{Maraschi92,Sikora94,ssc1}, while the hadronic mechanism produces higher energy photons by proton-synchrotron emission or photo-meson production~\cite{mannheim1,mannheim2,Muecke2003}. The values of $B^{\rm jet}_0$ and $y_{\rm em}$ are crucial for the photon-ALP system: the hadronic model requires higher $B^{\rm jet}_0$ and $y_{\rm em}$ as compared with the leptonic one. In addition, a higher initial degree of linear polarization $\Pi_{L,0}$ is predicted by hadronic models with respect to that expected in the leptonic scenario.

We compute the photon-ALP beam propagation in the jet comoving frame: thus, we must apply the transformation $E \to \gamma E$, where $\gamma$ is the Lorentz factor, when passing to the fixed frames of the regions to be considered below. The complete calculation and analysis of the photon-ALP conversion in the jet is reported in~\cite{trg2015}.

The parameter values concerning $B^{\rm jet}_0$, $y_{\rm em}$, $\gamma$ and $\Pi_{L,0}$ vary from leptonic to hadronic models, and also for the specific sources: we refer the reader to the subsequent Subsections about the specific BL~Lacs for more details.

\subsection{Host galaxy}

BL~Lacs are commonly hosted in elliptical galaxies, which are characterized by a turbulent magnetic field ${\bf B}_{\rm host}$, which is usually described by means of a domain-like model. Typical values of the strength and of the coherence length of ${\bf B}_{\rm host}$ are $B_{\rm host} \simeq 5 \, \mu{\rm G}$ and $L_{\rm dom}^{\rm host} \simeq 150 \, {\rm pc}$, respectively~\cite{moss1996}. Photon-ALP conversion is inefficient in this region because of the very high $\gamma \leftrightarrow a$ oscillation length if compared to $L_{\rm dom}^{\rm host}$ as shown in~\cite{trgb2012}.%, we accurately take it into account in our calculation. 

\subsection{Galaxy cluster}

All BL~Lacs considered in this paper are located outside rich galaxy clusters. In this paper, we are thus considering an alternate scenario with respect to that explored in~\cite{galantiPol}, where blazars were supposed to be hosted in rich galaxy clusters. %In that case photon-ALP conversion is efficient and has thus been instead properly considered. For a detailed description of the galaxy cluster properties, which are important for photon-ALP conversion, see~\cite{galantiPol,grtClu,grRew}.
In the following, we do not take into account photon-ALP interaction within this region, as photon-ALP conversion is not efficient.
%for the above-mentioned reasons.

\subsection{Extragalactic space}

The extragalactic space is a region of high photon absorption in the VHE band because of the existence of the extragalactic background light (EBL)~\cite{franceschinirodighiero,dgr2013,gptr}. However, we are dealing with much lower energies in the present paper, and so the EBL absorption turns out to be negligible. However, the photon-ALP interaction in this region modifies $P_{\gamma \to \gamma}$ and the corresponding $\Pi_L$ and $\chi$ in a way which depends on the strength and morphology of the extragalactic magnetic field ${\bf B}_{\rm ext}$.

Unfortunately, ${\bf B}_{\rm ext}$ is currently poor known: only the most recent bound $10^{- 7} \, {\rm nG} \leq {B}_{\rm ext} \leq 1.7 \, {\rm nG}$ on the scale of ${\cal O} (1) \, {\rm Mpc}$ exists~\cite{neronov2010,durrerneronov,pshirkov2016}. As a matter of fact, ${\bf B}_{\rm ext}$ is commonly described by means of a domain-like model~\cite{kronberg1994,grassorubinstein}, wherein ${\bf B}_{\rm ext}$ changes discontinuously at the interface of two adjacent domains.  We employ the new improved version of this model -- described in~\cite{grSM} -- which avoids such an unphysical discontinuity of the original proposal which produces inaccurate results. In the new model, ${\bf B}_{\rm ext}$  possesses a constant strength and orientation in the central part of each domain of size $L_{\rm dom}^{\rm ext}$, but smoothly and continuously connects to the orientation of ${\bf B}_{\rm ext}$ in the adjacent domains~\cite{grSM}.

Since outflows from primeval galaxies predict rather high values of $B_{\rm ext}$ with $B_{\rm ext} = {\cal O}(1) \, \rm nG$ for $L_{\rm dom}^{\rm ext} = {\cal O}(1) \, \rm Mpc$~\cite{reessetti,hoyle,kronberg1999,furlanettoloeb}, we assume $B_{\rm ext} = 1 \, \rm nG$ for definiteness and $L_{\rm dom}^{\rm ext}$ randomly varying with a power-law distribution function $\propto (L_{\rm dom}^{\rm ext})^{-1.2}$ in the range $(0.2 -10) \, \rm Mpc$ and with $\langle L_{\rm dom}^{\rm ext} \rangle = 2 \, \rm Mpc$.

\subsection{Milky Way}

There exist nowadays detailed maps of the Milky Way electron number density $n_e^{\rm MW}$ and magnetic field ${\bf B}_{\rm MW}$. In particular, we employ the model developed in~\cite{ymw2017} concerning $n_e^{\rm MW}$ and that by Jansson and Farrar~\cite{jansonfarrar1,jansonfarrar2,BMWturb} regarding ${\bf B}_{\rm MW}$. Although also the ${\bf B}_{\rm MW}$ model by Pshirkov {\it et al.}~\cite{pshirkovMF2011} is present in the literature and leads to similar results, we prefer that by Jansson and Farrar~\cite{jansonfarrar1,jansonfarrar2,BMWturb}, since it turns out to be more complete, as it accurately describes also the Galactic halo component.

Although the regular component of ${\bf B}_{\rm MW}$ gives the maximal contribution to the photon-ALP interaction, we take into account also the turbulent part of ${\bf B}_{\rm MW}$ in the calculation of the photon-ALP conversion in the Milky Way. For more details see~\cite{gtre2019}.

\subsection{Overall photon-ALP beam propagation}

By combining in the correct order the transfer matrices of the photon-ALP beam propagating in the regions described above, we can evaluate the total transfer matrix of the system $\cal U$. From $\cal U$ we obtain $P_{\gamma \to \gamma}$ and the corresponding final $\Pi_L$ and $\chi$ for all the considered BL~Lacs, by employing the strategy developed in~\cite{galantiPol,grtClu,grRew}, to which we refer the reader for more details about the calculation.

\section{Results}

Before proceeding to present our results, we must deal with a technological problem, which may in principle limit the effectiveness of our findings. Since real polarimeters do not possess very high spatial resolution, we cannot distinguish photons coming from different zones inside the transverse section of the blazar jet. As a consequence, all photons are collected together and the polarization features may be washed out by the averaging procedure on all photons from the jet.

We must keep in mind that we observe emitted photons equally distributed on the jet section only in the case of a perfect alignment between the line of sight and the jet axis. This is an extreme situation. The other extreme option is represented by photons emitted perpendicularly to the jet axis {\it in the jet comoving frame}, with the resulting photons propagating along the external border of the jet (for the geometry of both the two extreme cases see Fig.~19 in~\cite{galantiPol}). The probability of being in one of the previous situations is obviously vanishingly small.

Realistically, we are in an intermediate situation, wherein photons propagate up to the Earth not along the jet axis and not close to the external conical border of the jet. Merely for statistical reasons, the region of close alignment is much less probable. Therefore, photons coming from a specific small zone in the jet transverse section are much more probable than the others (see~\cite{galantiPol} for a complete description of the problem and for the adopted strategy of evaluation). We take this fact into account by weighting the photons by means of a gaussian distribution. 
%In Fig.~20 of~\cite{galantiPol}, in the case of perfect alignment, we observe, as expected, a decrease of the final polarization features, but, in the general case of misalignment, we practically note no impact if compared to the other figures of the paper. 
Since the intermediate case is the only interesting one for real applications and represents what occurs in practice, we shall be concerned with such a situation only. In order to be conservative, we use the same magnetic field profile as reported in Eq.~(\ref{Bjet}) and a propagation distance in the jet of $1 \, \rm pc$.

We are now in a position to present our results about the final photon survival probability $P_{\gamma \to \gamma}$ and the corresponding photon degree of linear polarization $\Pi_L$ and polarization angle $\chi$, when photons are emitted at the jet base of the BL~Lacs: OJ~287, BL~Lacertae, Markarian~501 and 1ES~0229+200, and oscillate into ALPs in all the magnetized regions crossed by the photon-ALP beam (blazar jet, host galaxy, extragalactic space and Milky Way, see Sec. III). Since the real nature of the extragalactic magnetic field ${\bf B}_{\rm ext}$ and of other crossed turbulent magnetic fields is unknown, we evaluate several realizations of the photon-ALP beam propagation process by varying the magnetic field parameters and we calculate the probability density function $f_{\Pi}$ associated with the final $\Pi_L$ for the different realizations.

\begin{figure}
\centering
\includegraphics[width=0.46\textwidth]{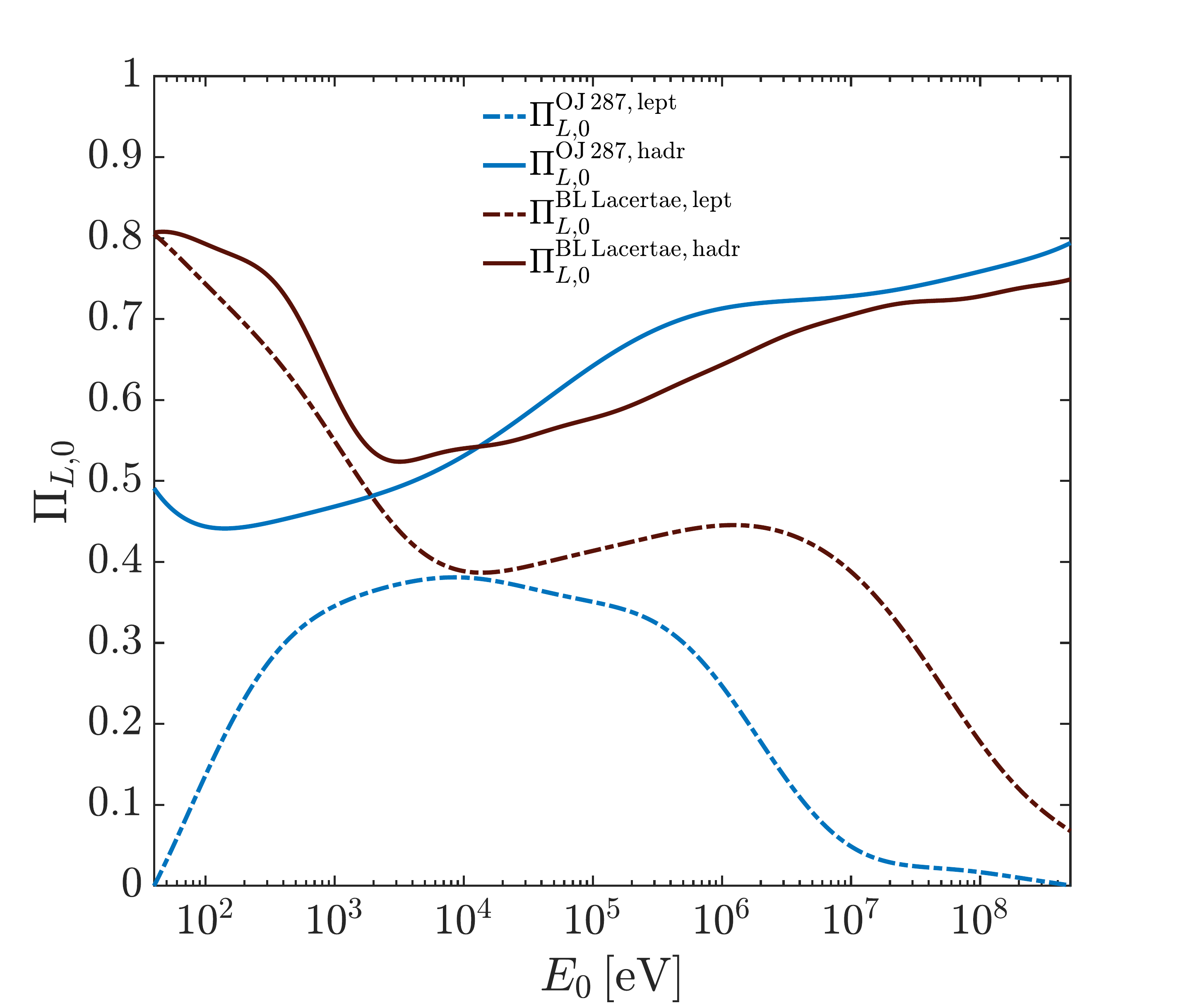}
\caption{\label{Polin} Initial degree of linear polarization $\Pi_{L,0}$ for the blazars OJ~287 and BL~Lacertae in the case of both leptonic and hadronic emission mechanisms, as derived in~\cite{zhangBott}.}
\end{figure}

We take ALP parameters within current bounds: $g_{a\gamma\gamma}=0.5 \times 10^{-11} \, \rm GeV^{-1}$ and the two ALP mass values: (i) $m_a \lesssim 10^{-14} \, \rm eV$, (ii) $m_a = 10^{-10} \, \rm eV$. We consider photons emitted at redshift $z$ with energy $E$ and with observed energy $E_0=E/(1+z)$ in the two ranges: (i) UV-X-ray band ($4 \times 10^{-2}\, {\rm keV} -10^2 \, \rm keV$), (ii) HE band $(10^{-1}\, {\rm MeV} -5 \times 10^2 \, \rm MeV)$. In Fig.~\ref{Polin} we report the initial degree of linear polarization $\Pi_{L,0}$ for the BL~Lacs OJ~287 and BL~Lacertae both in the leptonic and the hadronic scenarios, as derived in~\cite{zhangBott}. Instead, we take $\Pi_{L,0}=0.3$ in the UV-X-ray band for Markarian~501 in the leptonic model and the same value also for 1ES~0229+200 in both the leptonic and the hadronic cases (more about this, below).

In the following figures concerning the final $\Pi_L$ and $\chi$ for all considered BL~Lacs, we show also binned data of our results, in order to figure out how a possible detection of ALP-induced features may appear. The binning procedure takes the energy resolution of current polarimeters into account, which is expected to be worse with respect to spectrum-measuring observatories in the X-ray band~\cite{swift} by a factor $4 - 5$. In the HE range the energy resolution of spectral and polarization measurements is expected to be similar, as they derive from the same data~\cite{eastrogam1,eastrogam2,amego}. Correspondingly, we take $15 - 20$ bins per decade in the X-ray band and, conservatively, $8 - 10$ bins per decade in the HE range (see also note~\cite{footnote1} concerning the binning procedure).

\subsection{OJ~287}

OJ~287 is a low-frequency peaked blazar (LBL) observed at redshift $z=0.3056$. It is considered a good observational target for polarimetric studies both in the X-ray and HE range because of the high flux in both these energy bands~\cite{zhangBott}. We consider both a leptonic and hadronic emission model: correspondingly, we assume the typical LBL parameter values $B^{\rm jet}_0 = 1 \, \rm G$, $y_{\rm em} = 3 \times 10^{16} \, \rm cm$ and $\gamma = 10$ for the leptonic case and $B^{\rm jet}_0 = 20 \, \rm G$, $y_{\rm em} = 10^{17} \, \rm cm$ and $\gamma = 15$ concerning the hadronic one~\cite{LeptHadrBott}. The assumed initial degree of linear polarization $\Pi_{L,0}$ for both leptonic and hadronic scenarios is reported in Fig.~\ref{Polin}.

\begin{figure*}
\centering
\includegraphics[width=0.867\textwidth]{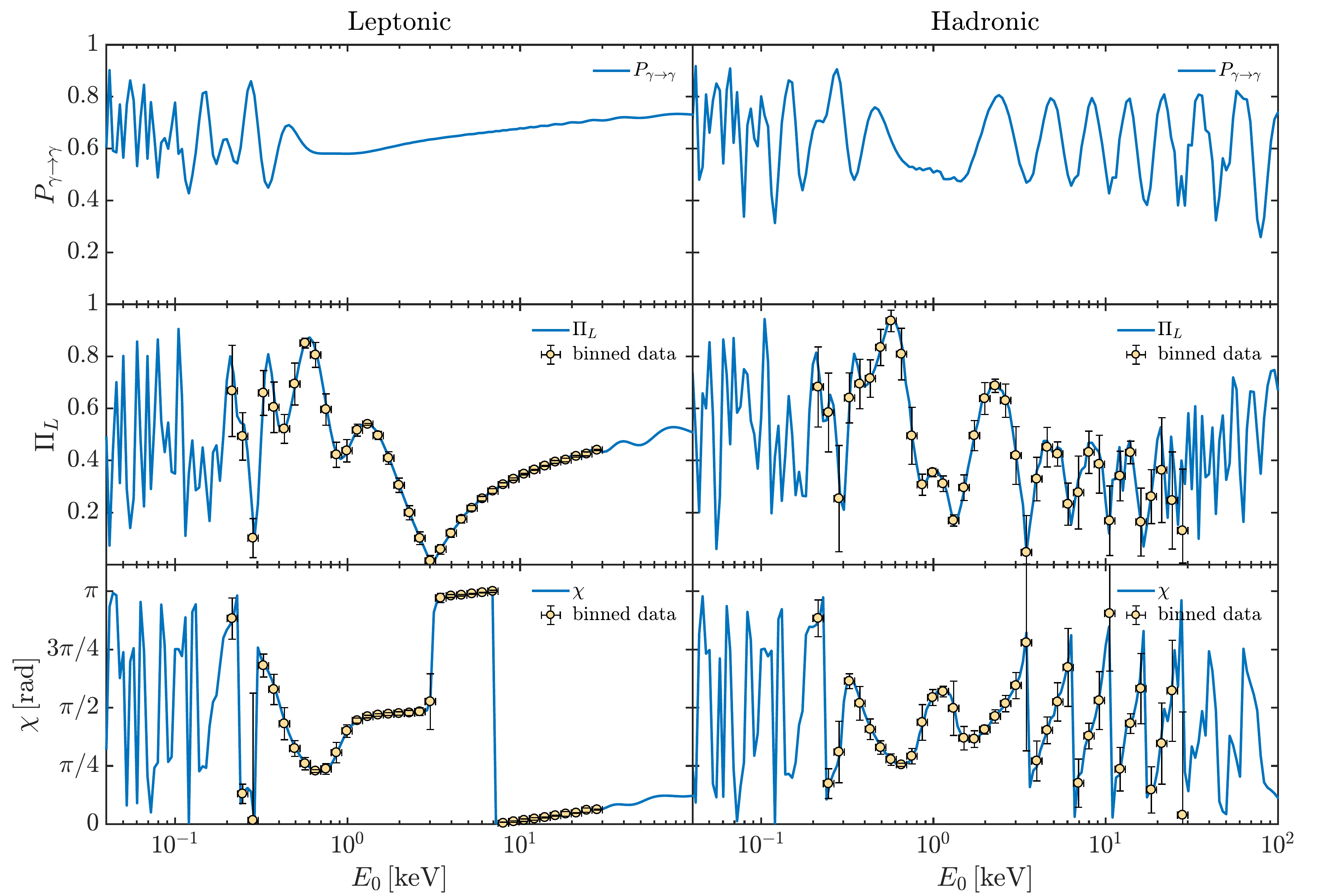}
\caption{\label{OJ287AllPolKeV} OJ~287: photon survival probability $P_{\gamma \to \gamma }$ (upper panels), corresponding final degree of linear polarization $\Pi_L$ (central panels) and final polarization angle $\chi$ (lower panels) in the energy range $(4 \times 10^{-2}-10^2) \, {\rm keV}$. We take $g_{a\gamma\gamma}=0.5 \times 10^{-11} \, \rm GeV^{-1}$, $m_a \lesssim 10^{-14} \, \rm eV$. We consider a leptonic and hadronic emission mechanism in the left and right column, respectively. Correspondingly, the initial degree of linear polarization is provided in Fig.~\ref{Polin}.}
\end{figure*}

\begin{figure*}
\centering
\includegraphics[width=0.867\textwidth]{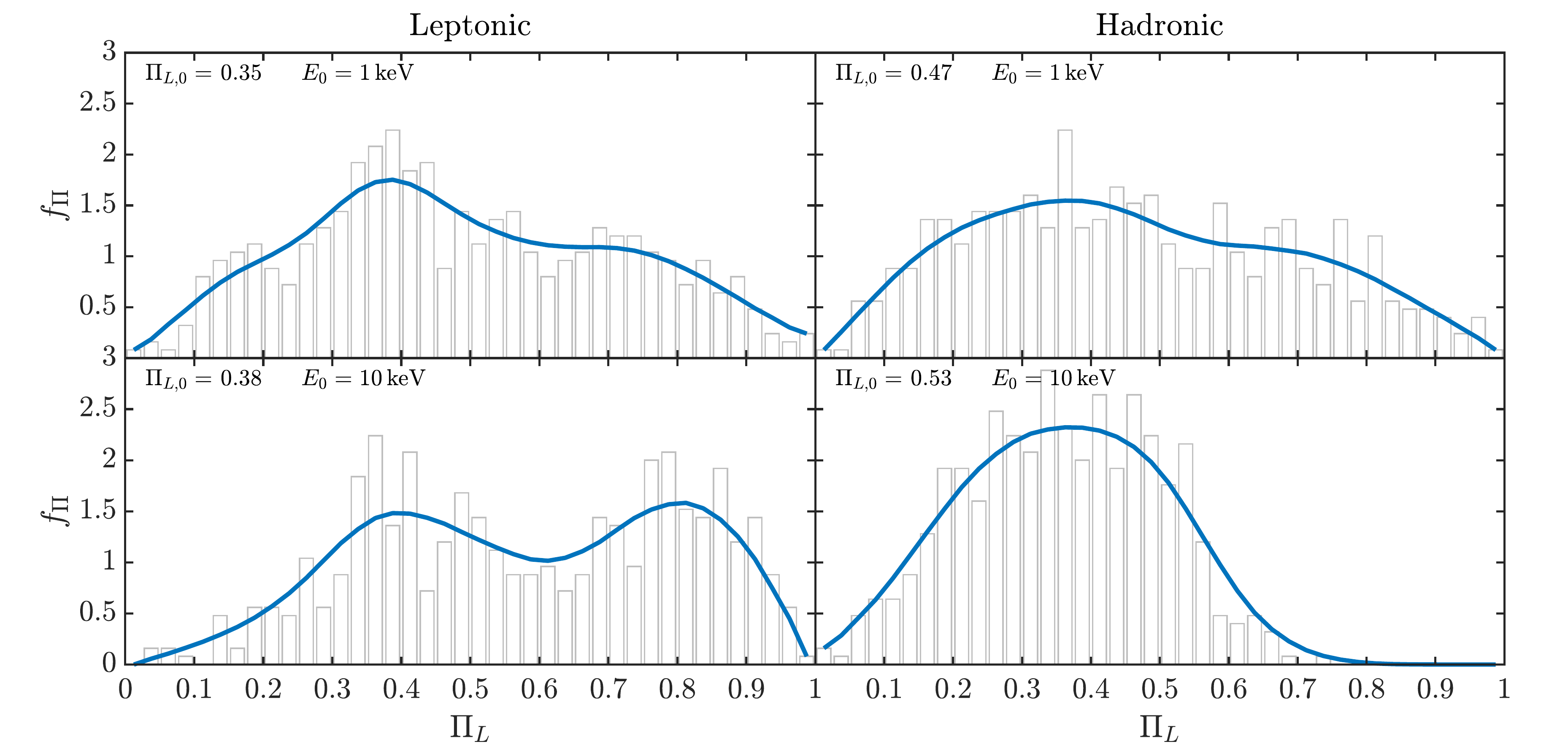}
\caption{\label{OJ287DensKeV} OJ~287: probability density function $f_{\Pi}$ arising from the plotted histogram for the final degree of linear polarization $\Pi_L$ at $1 \, \rm keV$ (upper panels) and $10 \, \rm keV$ (lower panels) by considering the system in Fig.~\ref{OJ287AllPolKeV}. We consider a leptonic and hadronic emission mechanism in the left and right column, respectively. Correspondingly, the initial degree of linear polarization $\Pi_{L,0}$ is provided in Fig.~\ref{Polin}.}
\end{figure*}

\begin{figure*}
\centering
\includegraphics[width=0.867\textwidth]{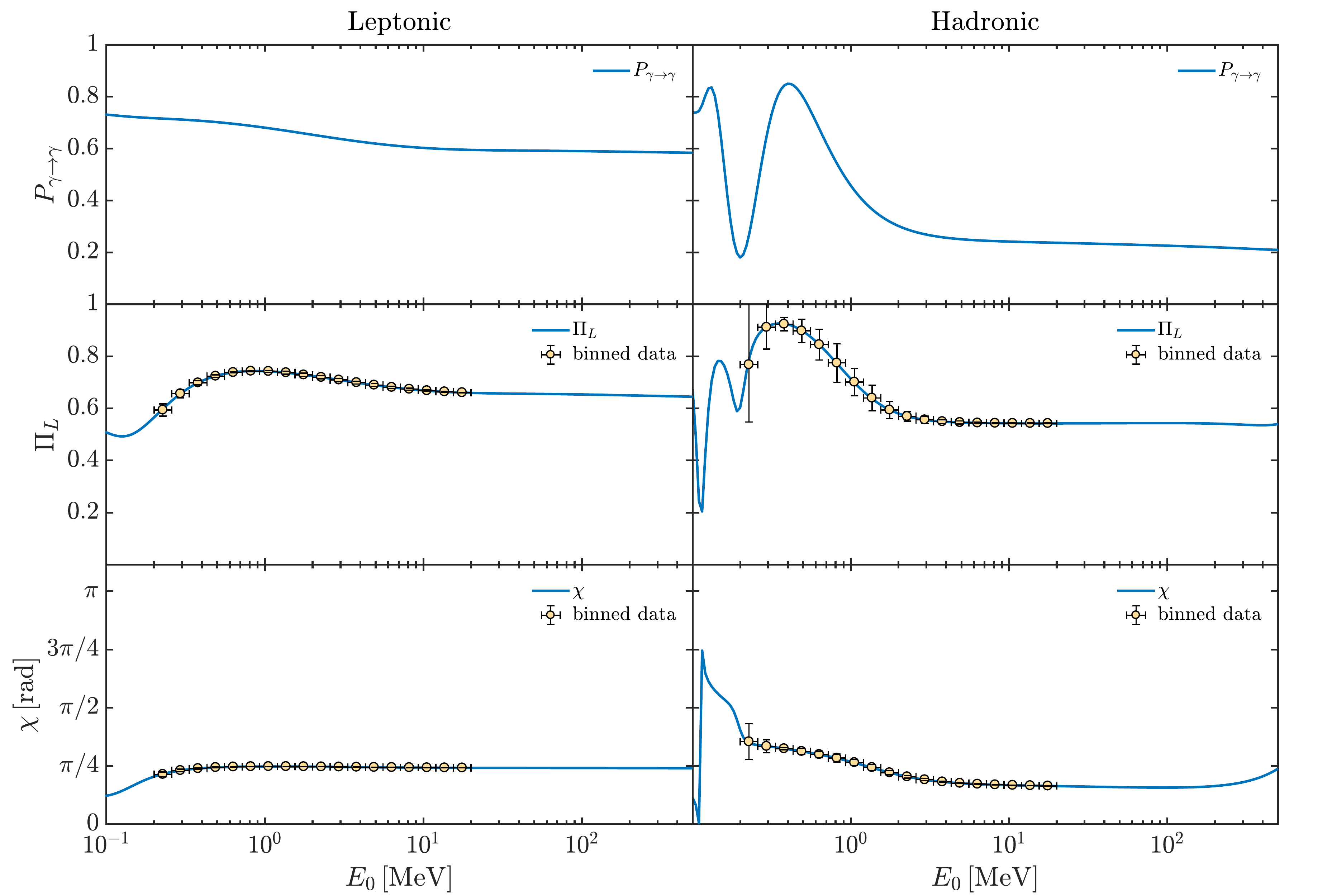}
\caption{\label{OJ287AllPolMeV} OJ~287: same as Fig.~\ref{OJ287AllPolKeV} but in the energy range $(10^{-1} -5 \times 10^2) \, {\rm MeV}$. We take $g_{a\gamma\gamma}=0.5 \times 10^{-11} \, \rm GeV^{-1}$, $m_a \lesssim 10^{-14} \, \rm eV$.}
\end{figure*}

\begin{figure*}
\centering
\includegraphics[width=0.867\textwidth]{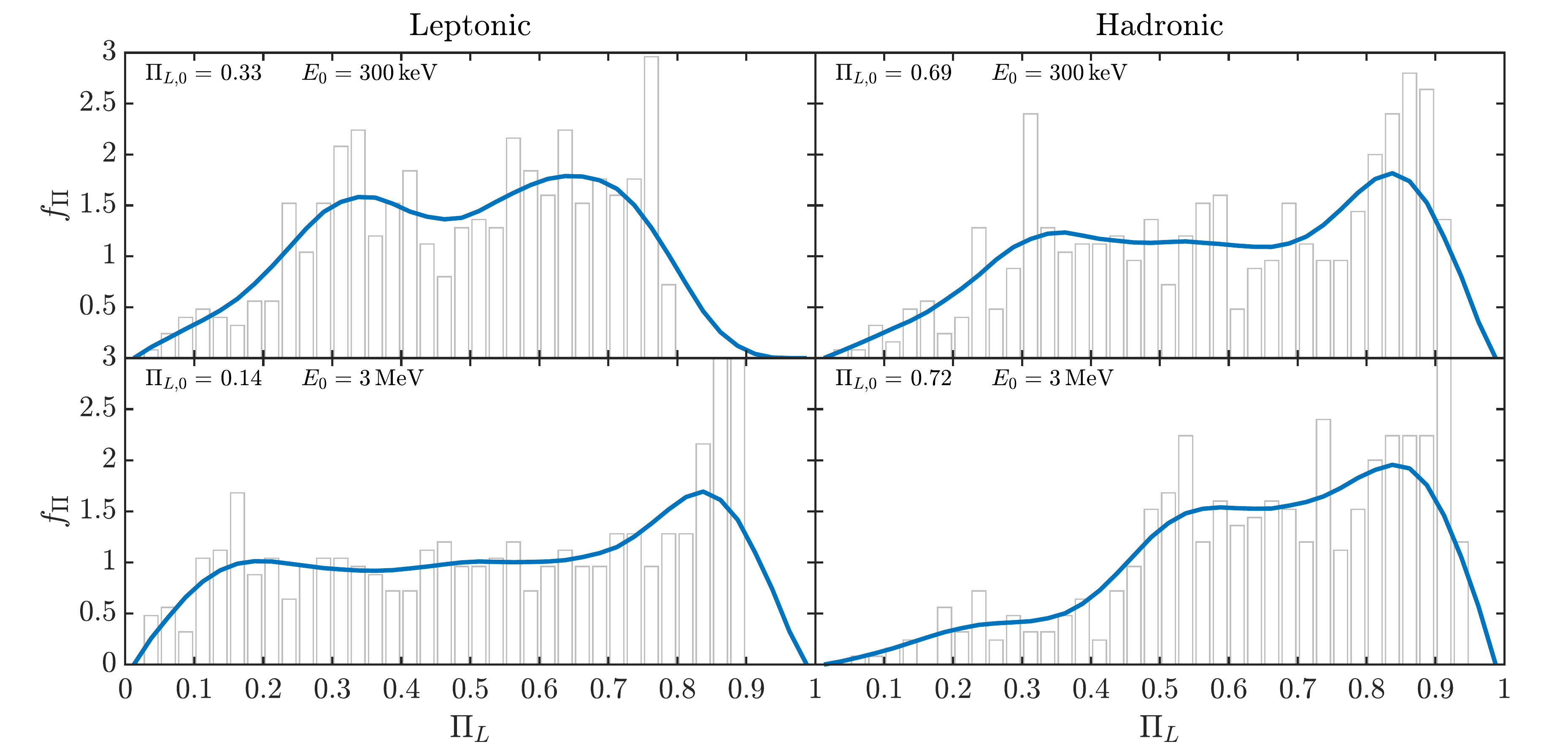}
\caption{\label{OJ287DensMeV} OJ~287: same as Fig.~\ref{OJ287DensKeV} but for the energies $300 \, \rm keV$ (upper panels) and $3 \, \rm MeV$ (lower panels) by considering the system in Fig.~\ref{OJ287AllPolMeV}.}
\end{figure*}

\begin{figure*}
\centering
\includegraphics[width=0.867\textwidth]{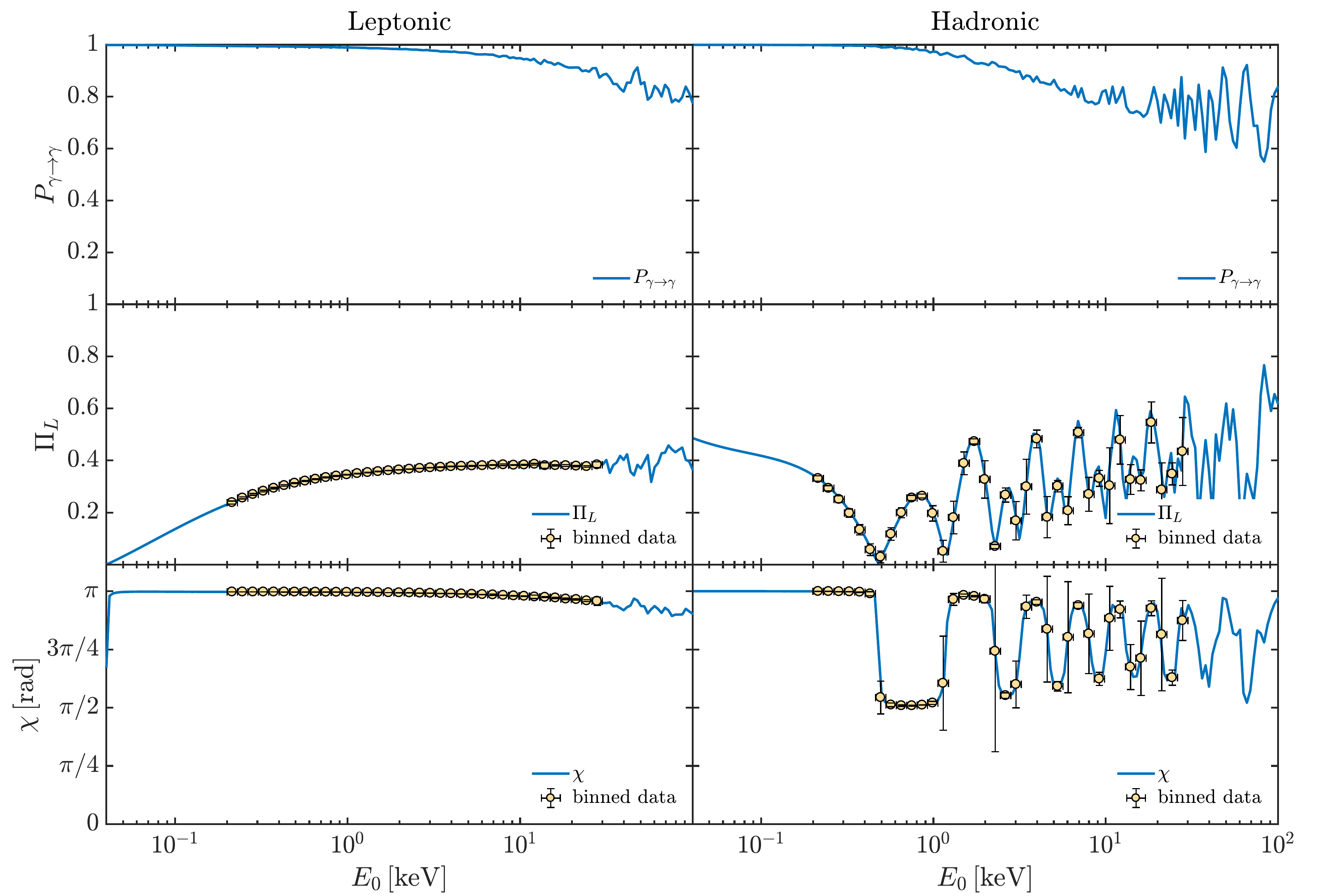}
\caption{\label{OJ287AllPolKeV-10} OJ~287: same as Fig.~\ref{OJ287AllPolKeV}. We take $g_{a\gamma\gamma}=0.5 \times 10^{-11} \, \rm GeV^{-1}$, $m_a = 10^{-10} \, \rm eV$.}
\end{figure*}

\begin{figure*}
\centering
\includegraphics[width=0.867\textwidth]{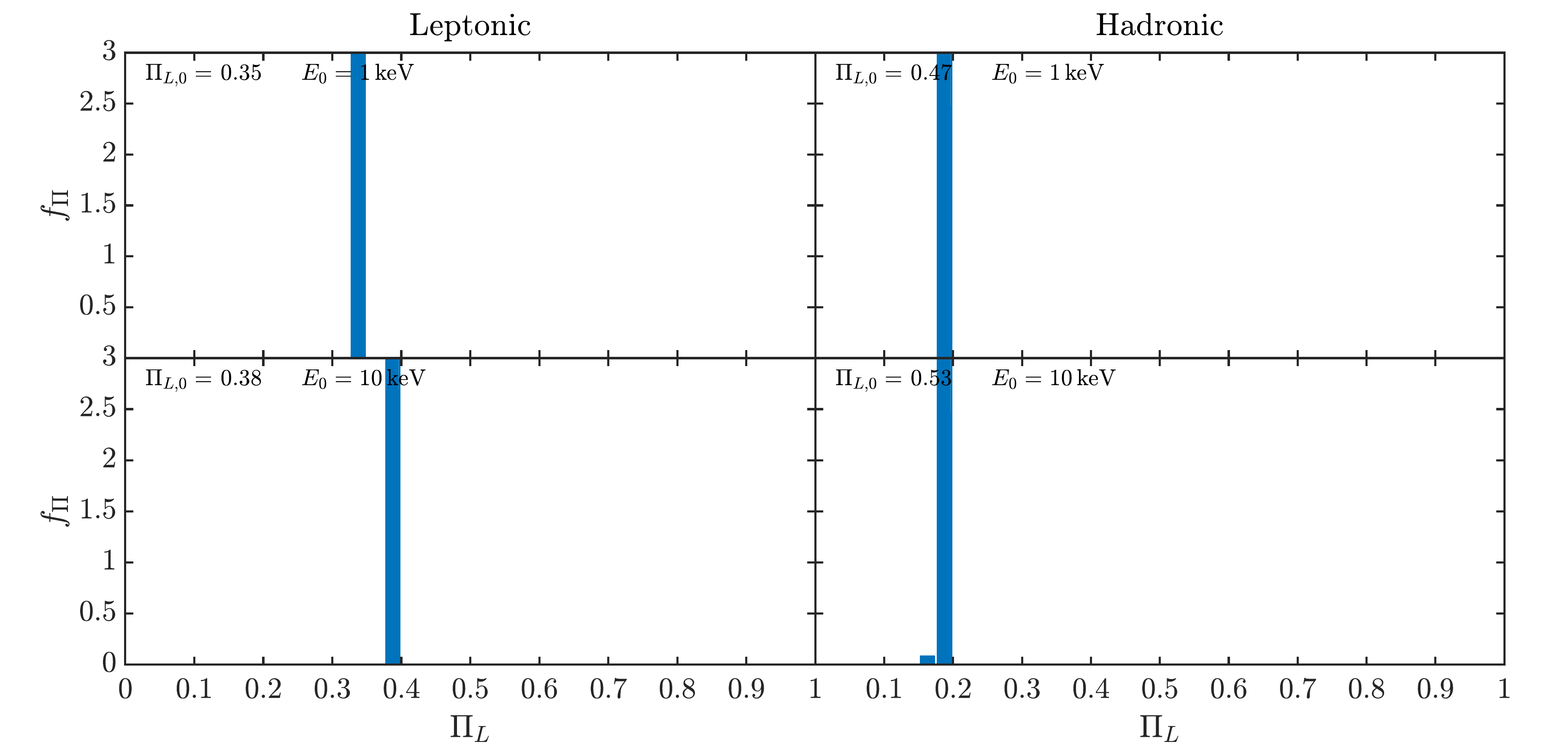}
\caption{\label{OJ287DensKeV-10} OJ~287: same as Fig.~\ref{OJ287DensKeV} by considering the system in Fig.~\ref{OJ287AllPolKeV-10}.}
\end{figure*}

\begin{figure*}
\centering
\includegraphics[width=0.867\textwidth]{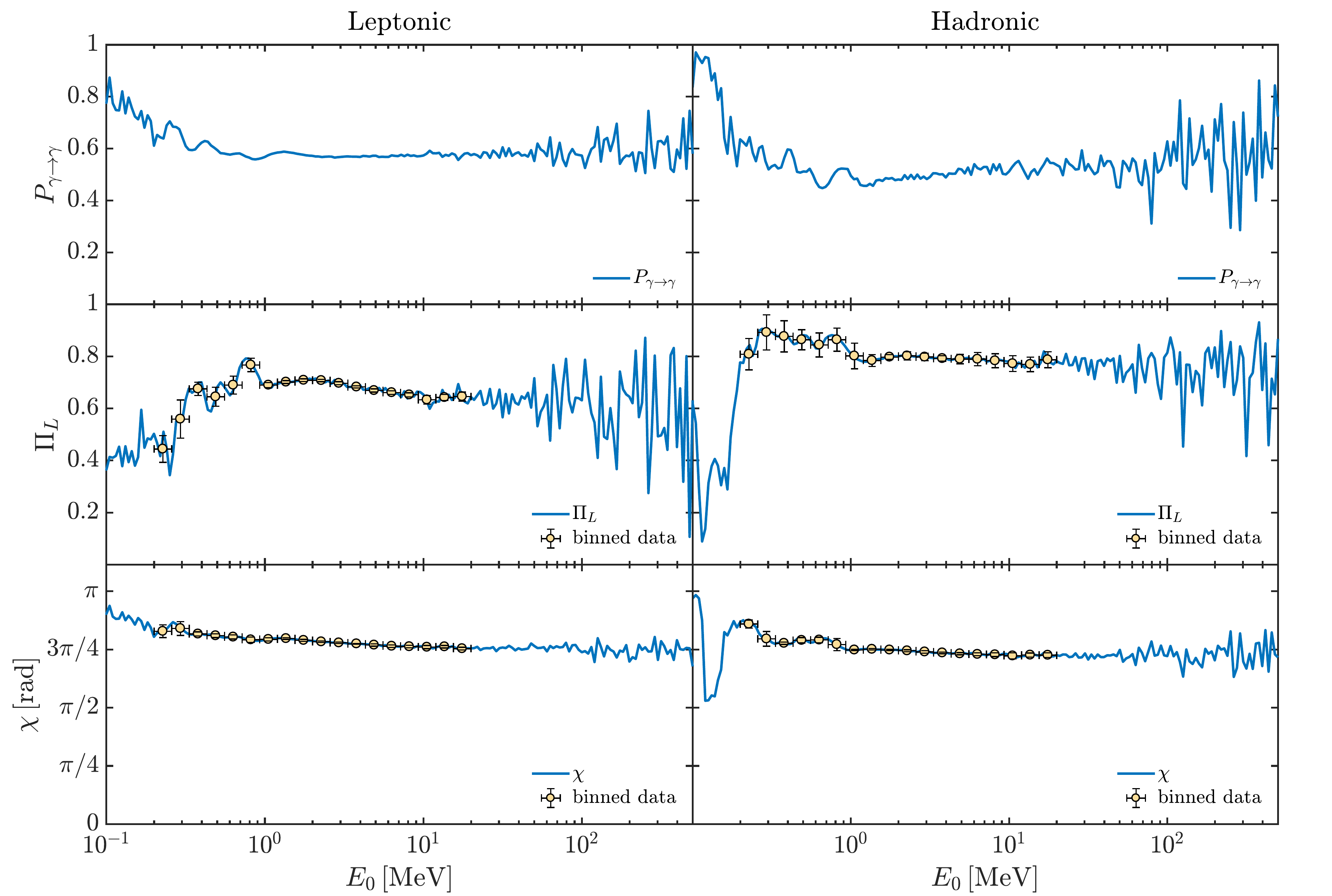}
\caption{\label{OJ287AllPolMeV-10} OJ~287: same as Fig.~\ref{OJ287AllPolMeV}. We take $g_{a\gamma\gamma}=0.5 \times 10^{-11} \, \rm GeV^{-1}$, $m_a = 10^{-10} \, \rm eV$.}
\end{figure*}

\begin{figure*}
\centering
\includegraphics[width=0.867\textwidth]{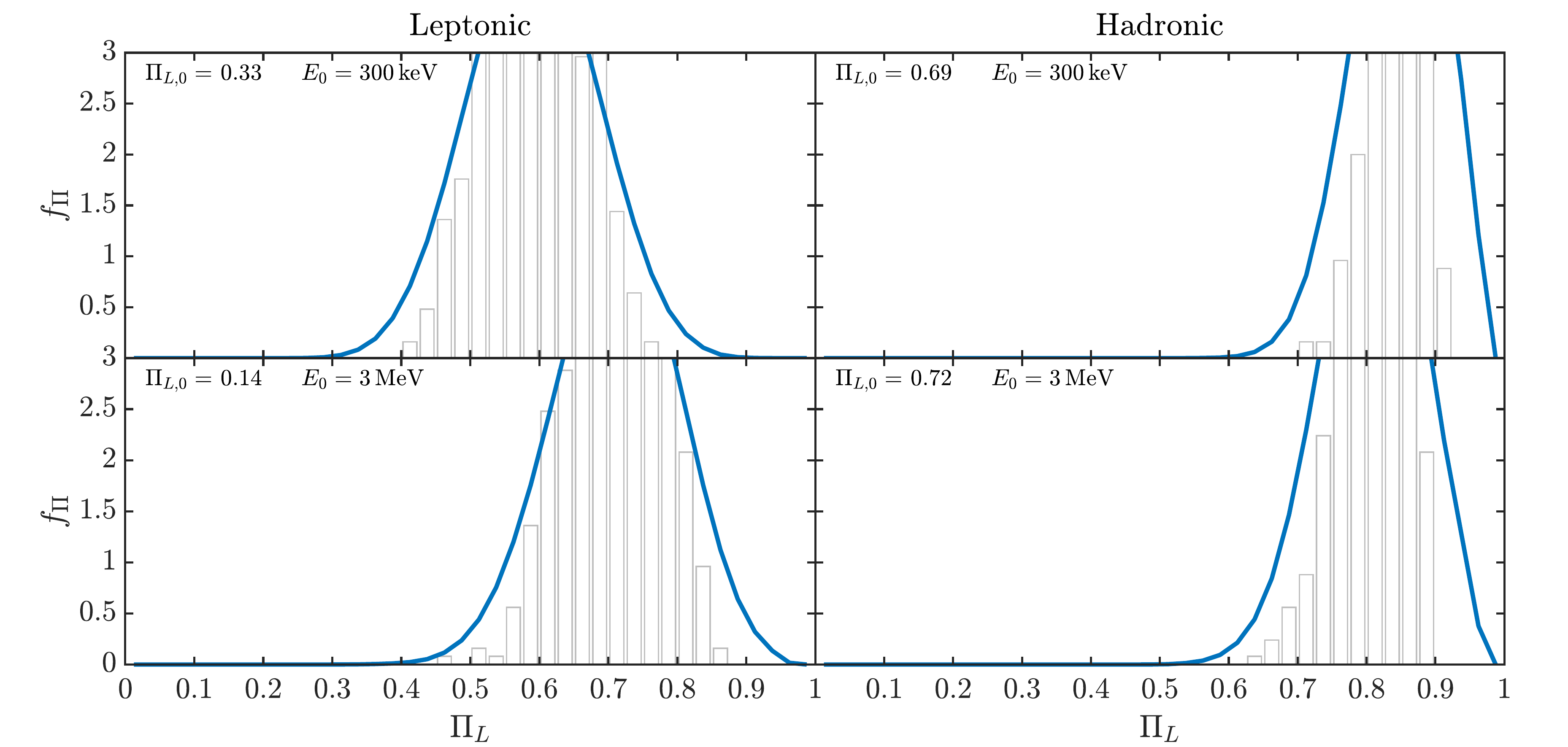}
\caption{\label{OJ287DensMeV-10} OJ~287: same as Fig.~\ref{OJ287DensMeV} by considering the system in Fig.~\ref{OJ287AllPolMeV-10}.}
\end{figure*}

We start by considering the case of an ALP mass $m_a \lesssim 10^{-14} \, \rm eV$. Our results concerning the UV-X-ray band ($4 \times 10^{-2} \, {\rm keV}-10^2 \, \rm keV$) are shown in Figs.~\ref{OJ287AllPolKeV} and~\ref{OJ287DensKeV}. In particular, we report $P_{\gamma \to \gamma}$ in the top panel of Fig.~\ref{OJ287AllPolKeV} and the corresponding final $\Pi_L$ and $\chi$ in the central and lower panel of Fig.~\ref{OJ287AllPolKeV}, respectively. As the photon-ALP beam propagates in the weak mixing regime because of the significant contribution of the plasma term (see also~\cite{grSM}), Fig.~\ref{OJ287AllPolKeV} shows an energy dependence of $P_{\gamma \to \gamma}$ and of the corresponding final $\Pi_L$ which is strongly modified with respect to the initial $\Pi_{L,0}$ both in the leptonic and in the hadronic cases. As noted in~\cite{galantiPol}, the weak mixing regime is not limited to a small energy range but extends for several energy decades because of the high variation of ${B}^{\rm jet}$ and $n_e^{\rm jet}$ [see Eqs.~(\ref{Bjet}) and~(\ref{njet}), respectively]. This remark applies to all the following cases wherein the system, for different parameters values, lies in the weak mixing regime. The binned data in Fig.~\ref{OJ287AllPolKeV} indicate that the high variability of $\Pi_L$ in both the leptonic and the hadronic cases can be detected by IXPE~\cite{ixpe}, eXTP~\cite{extp}, XL-Calibur~\cite{xcalibur}, NGXP~\cite{ngxp} and XPP~\cite{xpp} for $E_0 \gtrsim 0.5 \, \rm keV$. As confirmed by the behavior of $\chi$ in Fig.~\ref{OJ287AllPolKeV}, we see that the hadronic case shows a higher energy variability as compared to the leptonic case because of the larger size of the jet region and of the corresponding value of ${B}^{\rm jet}_0$.

As the final $P_{\gamma \to \gamma}$ and the corresponding final $\Pi_L$ and $\chi$ depend on the particular choice of the orientation and coherence length of ${\bf B}_{\rm ext}$ (and of the other turbulent magnetic fields) but only the statistical properties of these quantities are known, our results in Fig.~\ref{OJ287AllPolKeV} are obtained for a particular realization of the photon-ALP beam propagation process. While the photon-ALP beam can physically experience only one realization at once, the propagation of the photon-ALP beam becomes a stochastic process and several realizations of it must be considered in order to infer its statistical properties and the robustness of our results. Consequently, we report the probability density function $f_{\Pi}$ for the final $\Pi_L$ of all realizations in Fig.~\ref{OJ287DensKeV}, where we consider the two benchmark energies $E_0 = 1 \, \rm keV$ and $E_0 = 10 \, \rm keV$. In all cases, we observe a broadening of the initial $\Pi_{L,0}$. For the hadronic case we generally find a decrease of the initial $\Pi_{L,0}$, while in the leptonic case and for $E_0 = 10 \, \rm keV$ the most probable result is $\Pi_L \gtrsim 0.8$, which is much higher than that predicted by conventional physics in both emission mechanisms. Thus, the detectability of the last feature appears robust. 

In the HE band -- we focus our attention on the range $(10^{-1} - 5 \times 10^2) \, \rm MeV$ -- and we follow the same strategy described above. Therefore, in both the leptonic and the hadronic cases we report $P_{\gamma \to \gamma}$, the related final 
$\Pi_L$ and $\chi$ for a particular realization of the photon-ALP beam propagation process in Fig.~\ref{OJ287AllPolMeV}, while the associated statistical properties obtained by considering all different realizations are shown in Fig.~\ref{OJ287DensMeV}, where we plot the corresponding $f_{\Pi}$. Since in both the leptonic and the hadronic scenarios the photon-ALP mixing term dominates over all the other effects (see also~\cite{grSM}) in almost the whole HE band, we see in Fig.~\ref{OJ287AllPolMeV} that the photon-ALP beam propagates in the strong mixing regime, so that the resulting binned data are almost energy independent, presenting less uncertainties than in the UV-X-ray band. As a result,  OJ~287 appears as a good target for observatories such as COSI~\cite{cosi}, e-ASTROGAM~\cite{eastrogam1,eastrogam2} and AMEGO~\cite{amego}. In Fig.~\ref{OJ287DensMeV} the behavior of $f_{\Pi}$ shows that in both the leptonic and the hadronic cases and for both the considered energies $E_0 = 300 \, \rm keV$ and $E_0 = 3 \, \rm MeV$ the final $\Pi_L$ is broadened. The case $E_0 = 3 \, \rm MeV$ appears as the most promising one both in the leptonic and the hadronic scenarios, since the most probable value for the final $\Pi_L$ is $\Pi_L \gtrsim 0.8$, which is higher than that predicted by conventional physics in both emission mechanisms.

We now consider the case of an ALP mass $m_a = 10^{-10} \, \rm eV$ by proceeding as above. In the UV-X-ray band the ALP mass term is so strong that in the leptonic scenario dominates in all regions crossed by the photon-ALP beam. Consequently, the final $\Pi_L$ is not modified since the photon-ALP conversion is totally inefficient, as shown by the left panels of Fig.~\ref{OJ287AllPolKeV-10} and by the behavior of $f_{\Pi}$ in the left panels of Fig.~\ref{OJ287DensKeV-10}. Instead, in the hadronic case, the photon-ALP conversion is sizable but only in the jet for the higher central vale of $B^{\rm jet}_0 = 20 \, \rm G$ with a resulting dimming of the initial $\Pi_{L,0}$ (see the behavior of $\Pi_L$ in the right panels of Fig.~\ref{OJ287AllPolKeV-10} and that of $f_{\Pi}$ in the right panels of Fig.~\ref{OJ287DensKeV-10}). In the present situation we have no substantial broadening in the final values of $\Pi_L$ for both $E_0 = 1 \, \rm keV$ and $E_0 = 10 \, \rm keV$, as shown in the right panels of Fig.~\ref{OJ287DensKeV-10}, since the photon-ALP conversion in the extragalactic space -- which is the maximal responsible for the broadening in the other cases -- is totally negligible.

In the HE band, the photon-ALP beam propagates in the weak mixing regime both in the leptonic and hadronic scenario, so that $P_{\gamma \to \gamma}$ and the corresponding final $\Pi_L$ and $\chi$ turn out to be energy dependent as shown in Fig.~\ref{OJ287AllPolMeV-10}. As a general trend, we observe an increase of the initial $\Pi_{L,0}$. We infer from the binned data that observatories like COSI~\cite{cosi}, e-ASTROGAM~\cite{eastrogam1,eastrogam2} and AMEGO~\cite{amego} can detect the above features for $E_0 \gtrsim (0.2 - 2) \, \rm MeV$. The behavior of $f_{\Pi}$ in Fig.~\ref{OJ287DensMeV-10} confirms that the most probable final values of $\Pi_L$ are an increase of the initial $\Pi_{L,0}$ with a moderate broadening for both the two considered benchmark energies $E_0 = 300 \, \rm keV$ and $E_0 = 3 \, \rm MeV$.

%\newpage

\subsection{BL~Lacertae}

BL~Lacertae is the prototype of the AGN class called BL~Lacertae object (BL~Lac). In particular, it is an intermediate-frequency peaked blazar (IBL) and has been observed at redshift $z=0.069$. Due to its relative proximity to the Earth and to its high flux in both the UV-X-ray and HE ranges, BL~Lacertae is regarded as a good observational target for polarimetric studies in both the bands~\cite{zhangBott}. Similarly to OJ~287, we take both a leptonic and a hadronic emission model into account. We take the following typical IBL parameter values: $B^{\rm jet}_0 = 1 \, \rm G$, $y_{\rm em} = 3 \times 10^{16} \, \rm cm$ and $\gamma = 10$ for the leptonic case and $B^{\rm jet}_0 = 20 \, \rm G$, $y_{\rm em} = 10^{17} \, \rm cm$ and $\gamma = 15$ concerning the hadronic one~\cite{LeptHadrBott}. For both the leptonic and the hadronic scenarios, the initial degree of linear polarization $\Pi_{L,0}$ is shown in Fig.~\ref{Polin}.

\begin{figure*}
\centering
\includegraphics[width=0.867\textwidth]{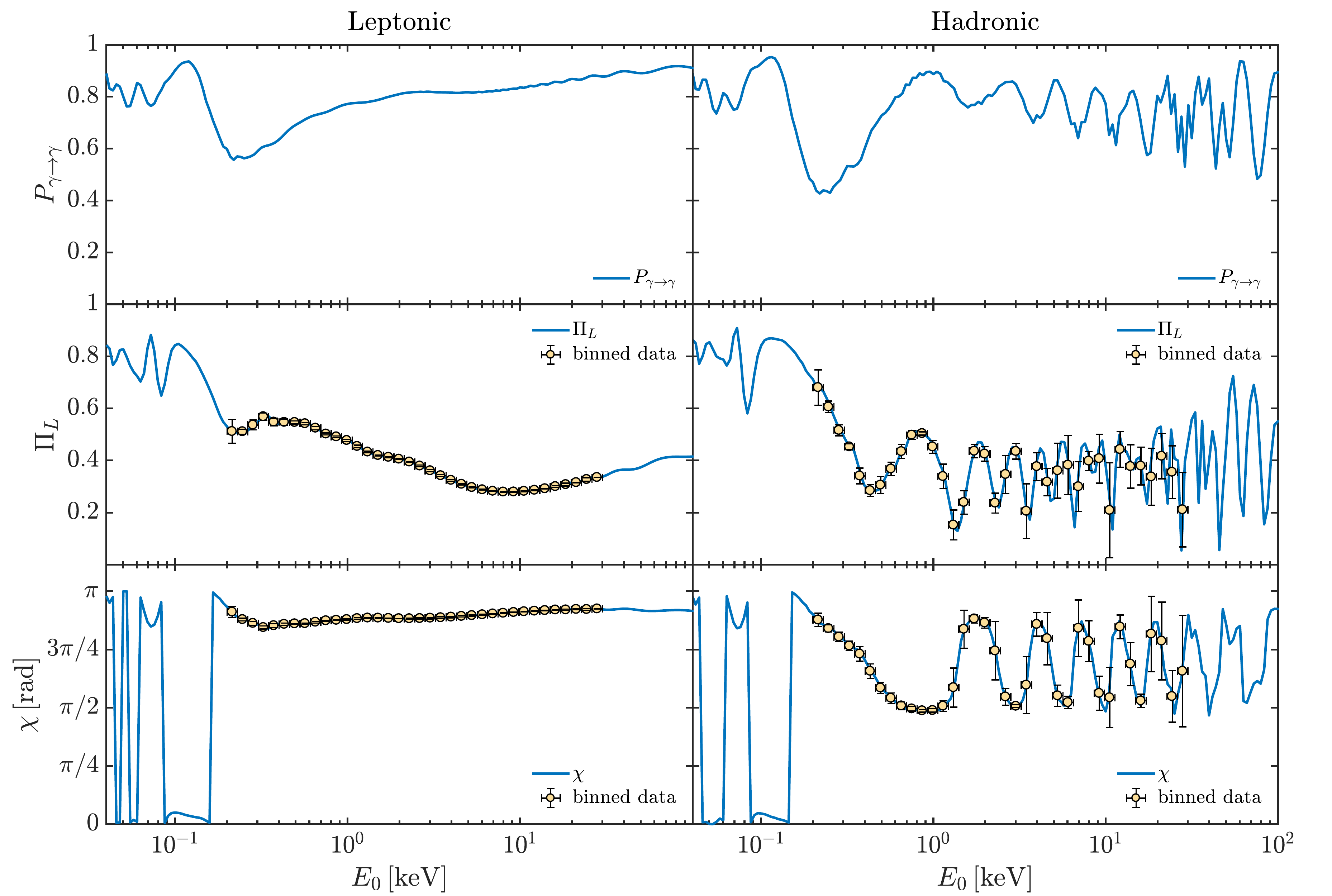}
\caption{\label{BLLACAllPolKeV} BL~Lacertae: photon survival probability $P_{\gamma \to \gamma }$ (upper panels), corresponding final degree of linear polarization $\Pi_L$ (central panels) and final polarization angle $\chi$ (lower panels) in the energy range $(4 \times 10^{-2}-10^2) \, {\rm keV}$. We take $g_{a\gamma\gamma}=0.5 \times 10^{-11} \, \rm GeV^{-1}$, $m_a \lesssim 10^{-14} \, \rm eV$. We consider a leptonic and hadronic emission mechanism in the left and right column, respectively. Correspondingly, the initial degree of linear polarization is provided in Fig.~\ref{Polin}.}
\end{figure*}

\begin{figure*}
\centering
\includegraphics[width=0.867\textwidth]{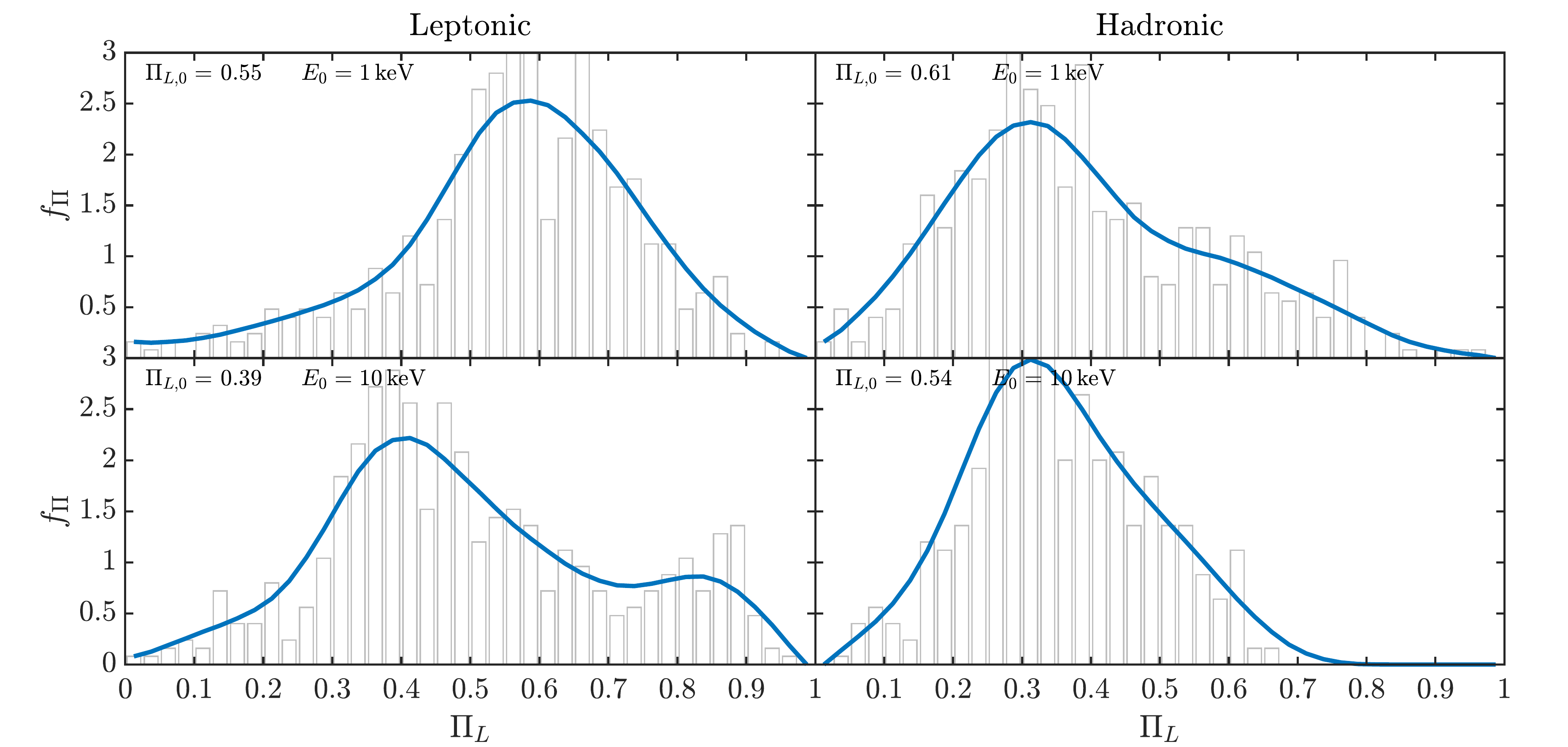}
\caption{\label{BLLACDensKeV} BL~Lacertae: probability density function $f_{\Pi}$ arising from the plotted histogram for the final degree of linear polarization $\Pi_L$ at $1 \, \rm keV$ (upper panels) and $10 \, \rm keV$ (lower panels) by considering the system in Fig.~\ref{BLLACAllPolKeV}. We consider a leptonic and hadronic emission mechanism in the left and right column, respectively. Correspondingly, the initial degree of linear polarization is provided in Fig.~\ref{Polin}.}
\end{figure*}

\begin{figure*}
\centering
\includegraphics[width=0.867\textwidth]{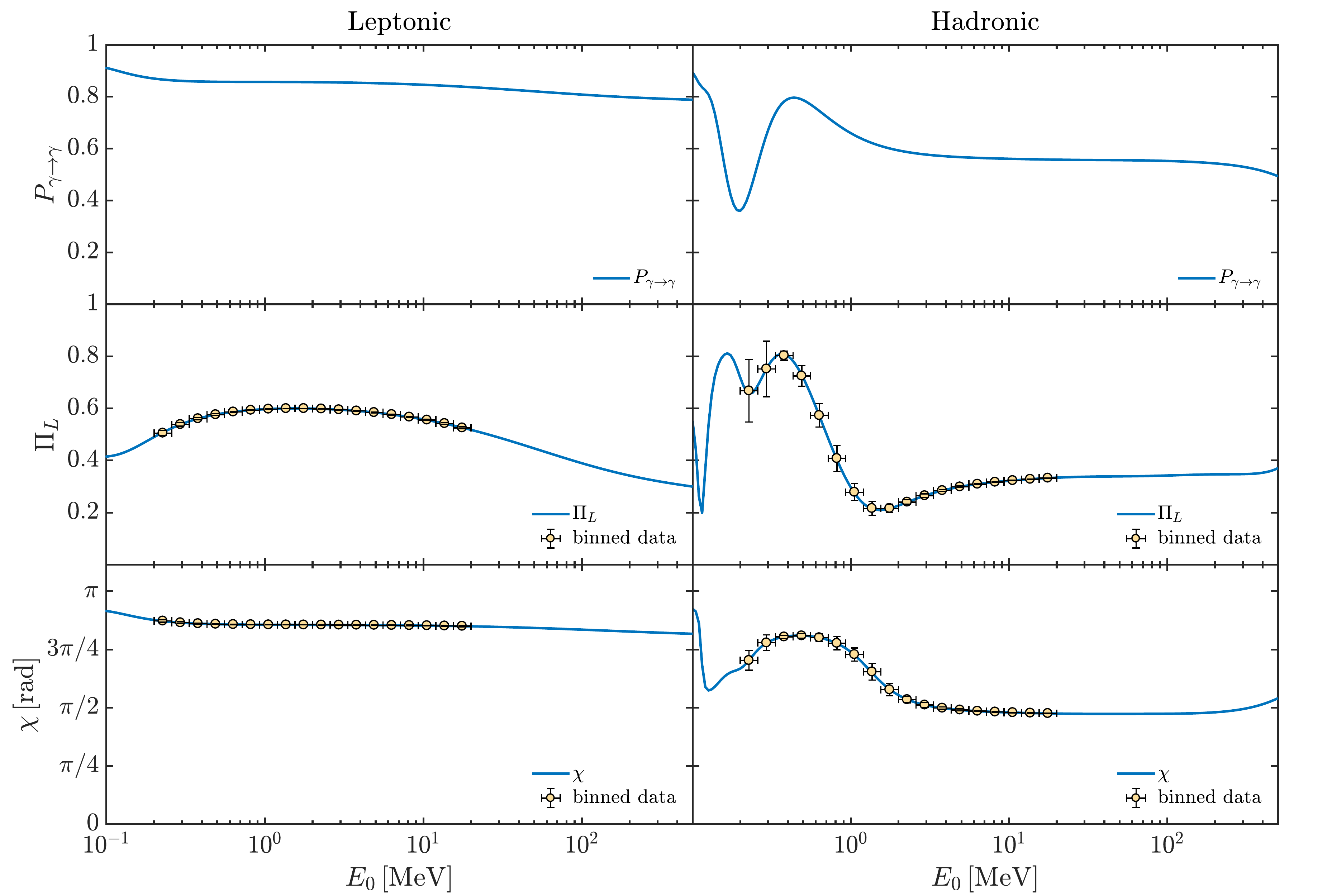}
\caption{\label{BLLACAllPolMeV} BL~Lacertae: same as Fig.~\ref{BLLACAllPolKeV} but in the energy range $(10^{-1} -5 \times 10^2) \, {\rm MeV}$. We take $g_{a\gamma\gamma}=0.5 \times 10^{-11} \, \rm GeV^{-1}$, $m_a \lesssim 10^{-14} \, \rm eV$.}
\end{figure*}

\begin{figure*}
\centering
\includegraphics[width=0.867\textwidth]{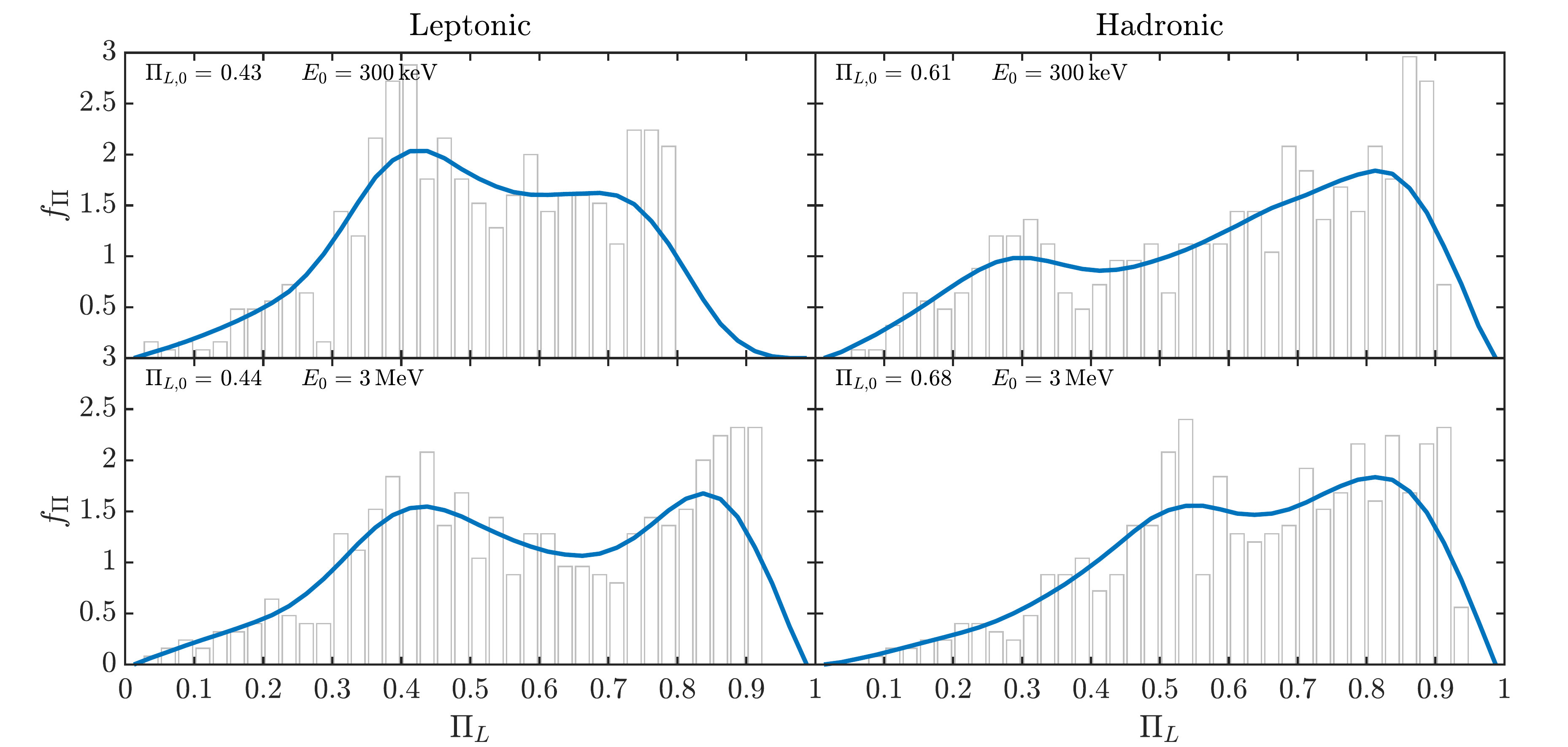}
\caption{\label{BLLACDensMeV} BL~Lacertae: same as Fig.~\ref{BLLACDensKeV} but for the energies $300 \, \rm keV$ (upper panels) and $3 \, \rm MeV$ (lower panels) by considering the system in Fig.~\ref{BLLACAllPolMeV}.}
\end{figure*}

\begin{figure*}
\centering
\includegraphics[width=0.867\textwidth]{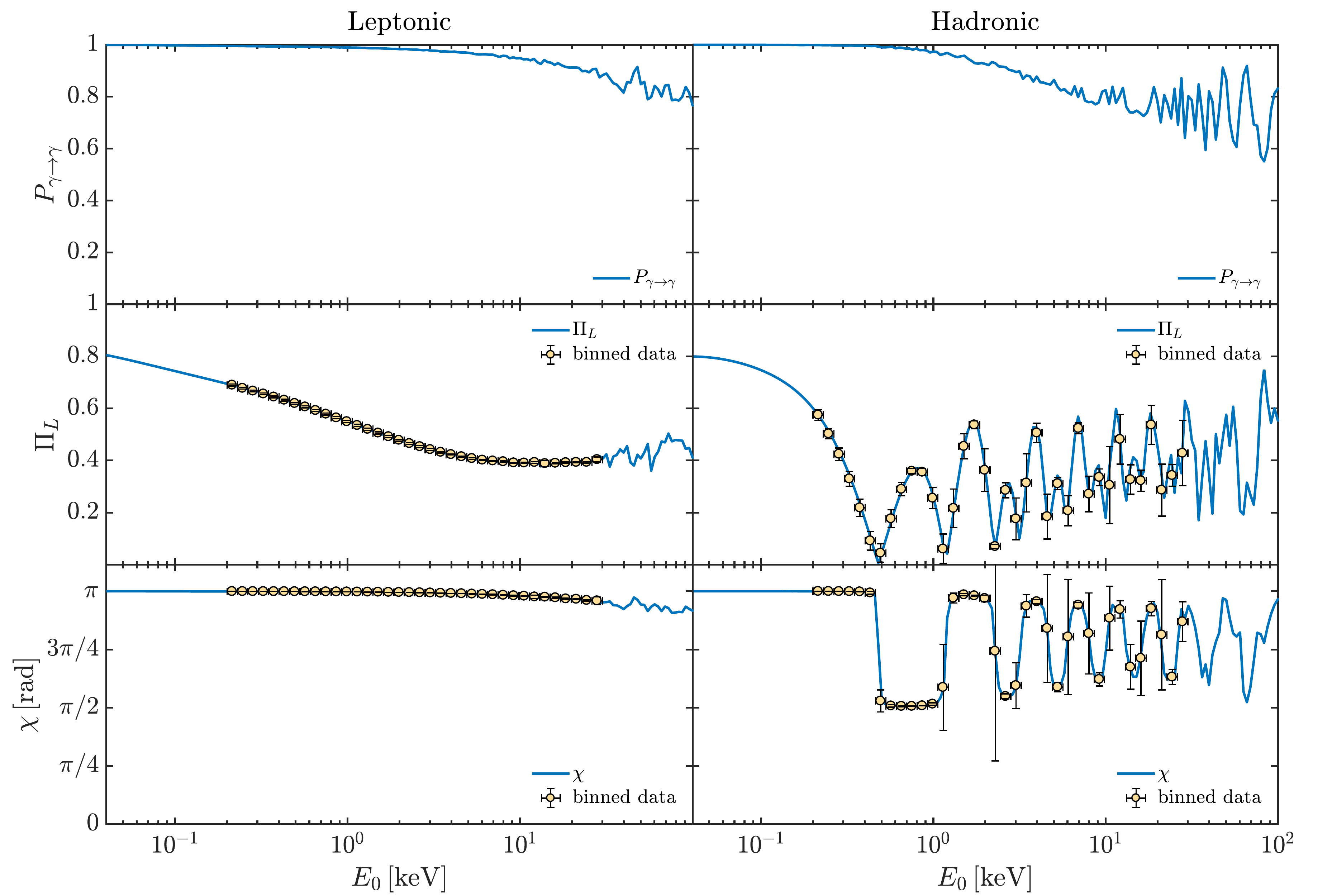}
\caption{\label{BLLACAllPolKeV-10} BL~Lacertae: same as Fig.~\ref{BLLACAllPolKeV}. We take $g_{a\gamma\gamma}=0.5 \times 10^{-11} \, \rm GeV^{-1}$, $m_a = 10^{-10} \, \rm eV$.}
\end{figure*}

\begin{figure*}
\centering
\includegraphics[width=0.867\textwidth]{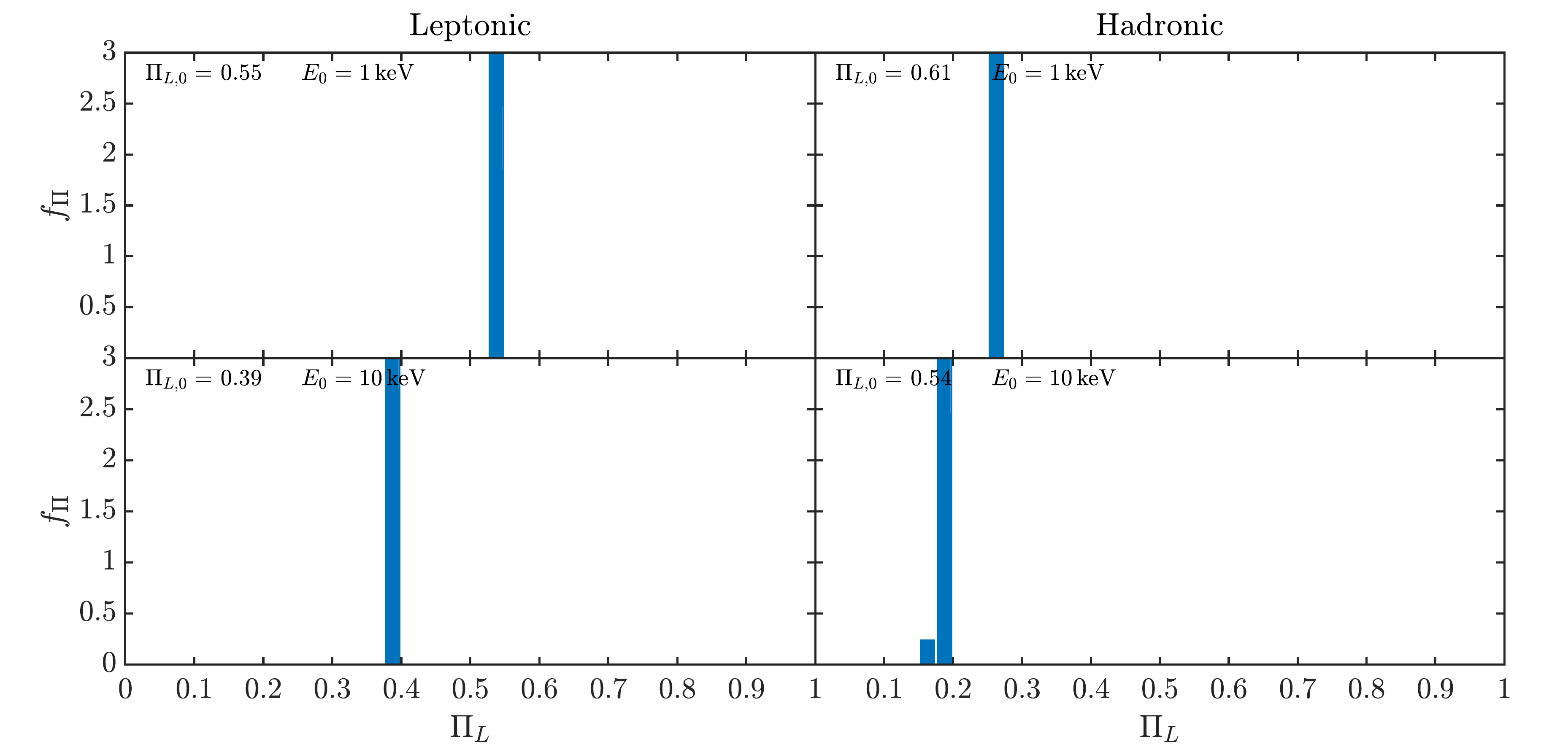}
\caption{\label{BLLACDensKeV-10} BL~Lacertae: same as Fig.~\ref{BLLACDensKeV} by considering the system in Fig.~\ref{BLLACAllPolKeV-10}.}
\end{figure*}

\begin{figure*}
\centering
\includegraphics[width=0.867\textwidth]{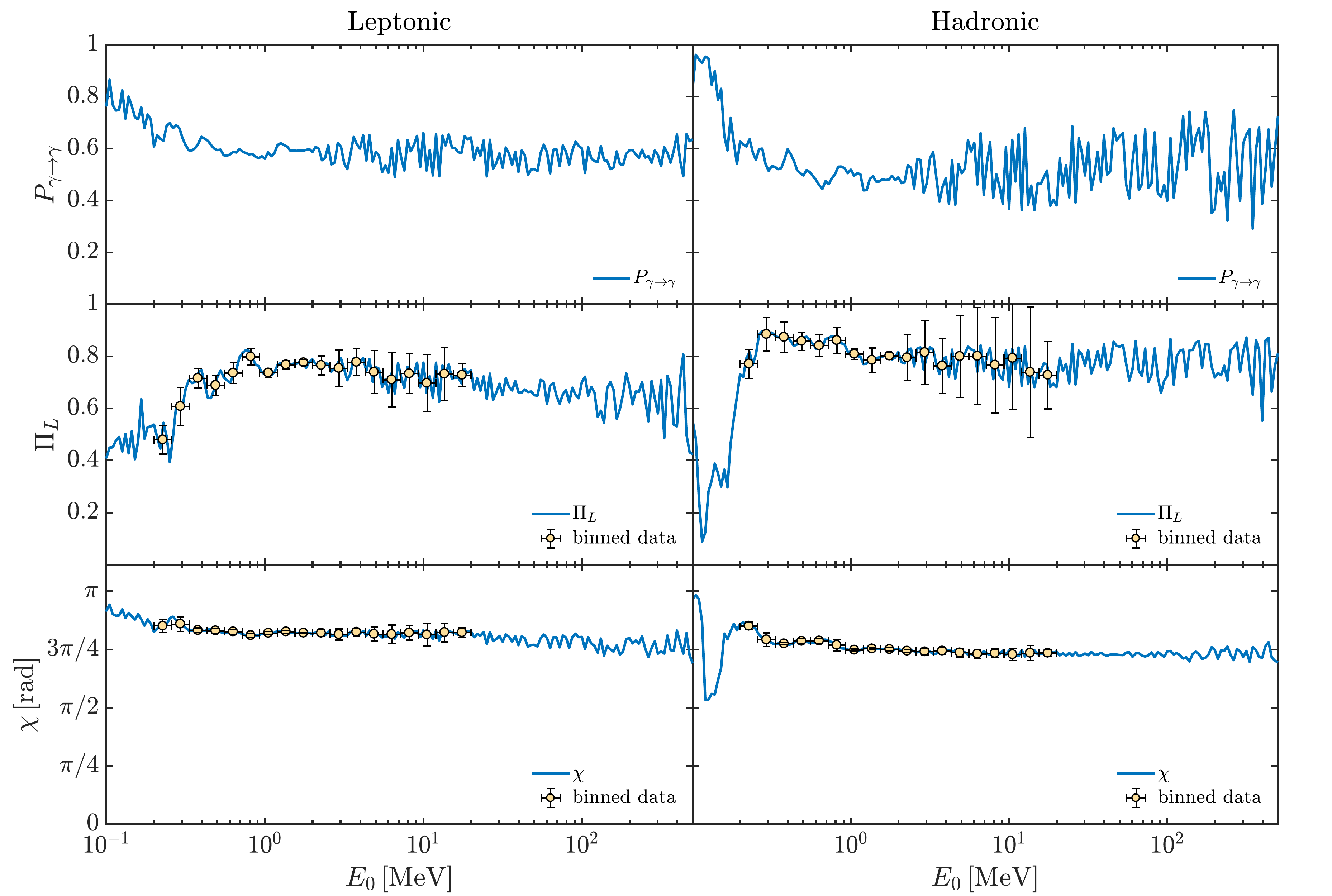}
\caption{\label{BLLACAllPolMeV-10} BL~Lacertae: same as Fig.~\ref{BLLACAllPolMeV}. We take $g_{a\gamma\gamma}=0.5 \times 10^{-11} \, \rm GeV^{-1}$, $m_a = 10^{-10} \, \rm eV$.}
\end{figure*}

\begin{figure*}
\centering
\includegraphics[width=0.867\textwidth]{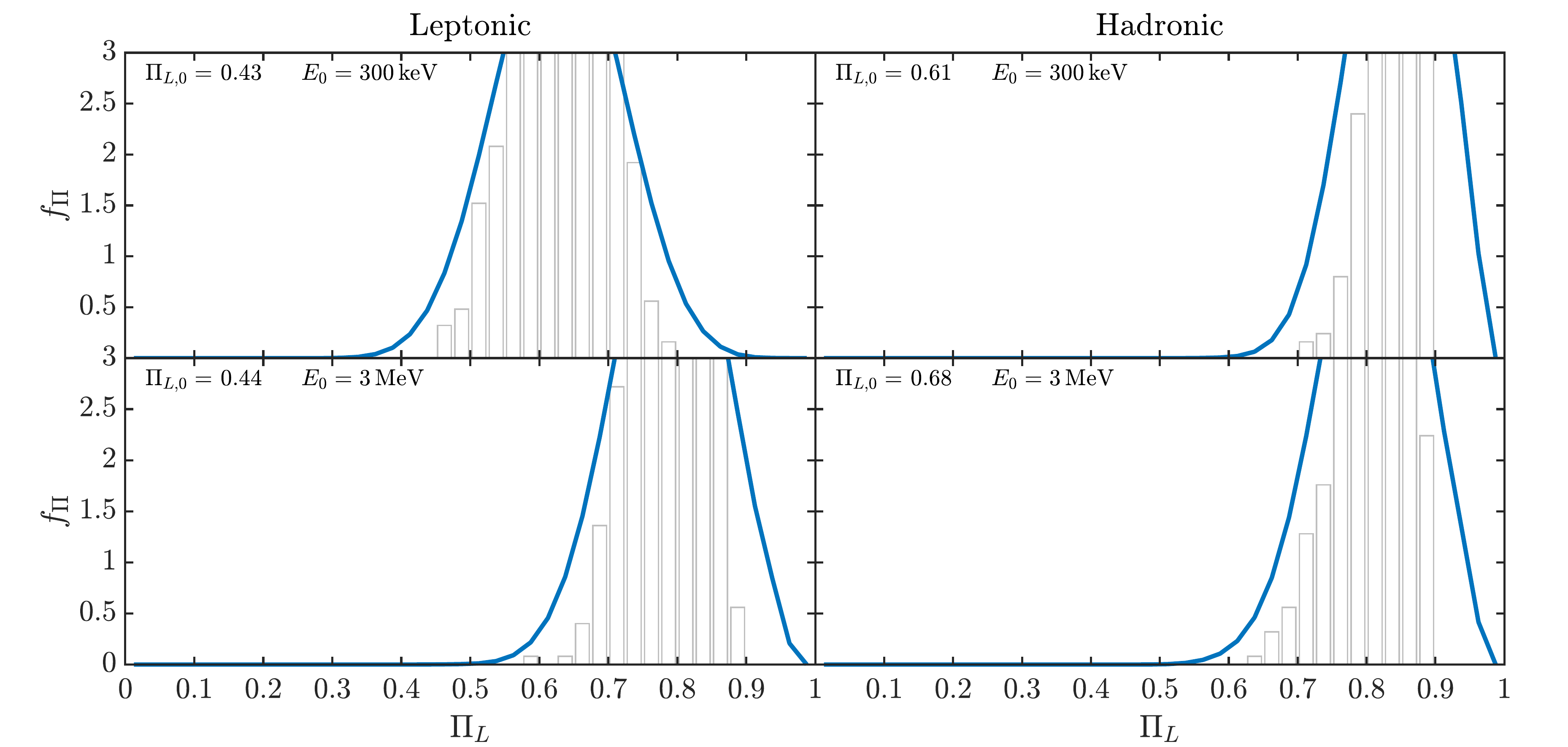}
\caption{\label{BLLACDensMeV-10} BL~Lacertae: same as Fig.~\ref{BLLACDensMeV} by considering the system in Fig.~\ref{BLLACAllPolMeV-10}.}
\end{figure*}

Since OJ~287 and BL~Lacertae possess similar features, they show comparable ALP-induced effects on the final photon polarization. This is the reason why we refer the reader to the previous Subsection about OJ~287 for a complete discussion about the results concerning BL~Lacertae, while here we will merely stress the few differences between the two sources.

In the case of an ALP mass $m_a \lesssim 10^{-14} \, \rm eV$, we exhibit our results concerning the UV-X-ray band in Figs.~\ref{BLLACAllPolKeV} and~\ref{BLLACDensKeV}, while our findings for the HE range are reported in Figs.~\ref{BLLACAllPolMeV} and~\ref{BLLACDensMeV}. In these cases, for BL~Lacertae we do not observe substantial modifications with respect to what we have found for OJ~287, apart from a less evident pseudo-oscillatory behavior of $P_{\gamma \to \gamma}$ and of the corresponding final $\Pi_L$ and $\chi$ especially in the UV-X-ray band, as reported in Fig.~\ref{BLLACAllPolKeV} (if compared to Fig.~\ref{OJ287AllPolKeV}) and a lower broadening of the final values of $\Pi_L$, as shown in Fig.~\ref{BLLACDensKeV} (see Fig.~\ref{OJ287DensKeV} for comparison). The reason for this behavior is that BL~Lacertae is closer to the Earth than OJ~287: thus, the photon-ALP conversion in the extragalactic space -- which is the major responsible for these effects -- is less effective. In the HE range, we do not find substantial modifications with respect to OJ~287 if we compare $\Pi_L$ in Fig.~\ref{BLLACAllPolMeV} with Fig.~\ref{OJ287AllPolMeV} and $f_{\Pi}$ in Fig.~\ref{BLLACDensMeV} with Fig.~\ref{OJ287DensMeV}.

In the case of an ALP mass $m_a = 10^{-10} \, \rm eV$, our results for the UV-X-ray band are reported in Figs.~\ref{BLLACAllPolKeV-10} and~\ref{BLLACDensKeV-10}, while our findings for the HE range are shown in Figs.~\ref{BLLACAllPolMeV-10} and~\ref{BLLACDensMeV-10}. In these cases the behavior of BL~Lacertae is totally similar to that of OJ~287, apart from the left panels of Figs.~\ref{BLLACAllPolKeV-10} and~\ref{BLLACDensKeV-10} concerning $\Pi_L$ and $f_{\Pi}$, respectively, if compared to the left panels of Figs.~\ref{OJ287AllPolKeV-10} and~\ref{OJ287DensKeV-10}. The reason for that lies in the fact that in the leptonic scenario and in the case $m_a = 10^{-10} \, \rm eV$, the photon-ALP conversion is negligible in the UV-X-ray band, so that the difference between BL~Lacertae and OJ~287 is only due to their distinct initial $\Pi_{L,0}$ (see Fig.~\ref{Polin}). 

Overall, also BL~Lacertae represents a good observational target for ALP studies with IXPE~\cite{ixpe}, eXTP~\cite{extp}, XL-Calibur~\cite{xcalibur}, NGXP~\cite{ngxp} and XPP~\cite{xpp} in the X-ray band and with COSI~\cite{cosi}, e-ASTROGAM~\cite{eastrogam1,eastrogam2} and AMEGO~\cite{amego} in the HE range especially in the cases discussed for OJ~287.

%\newpage

\subsection{Markarian~501}

Markarian~501 is a high-frequency peaked blazar (HBL) detected at redshift $z=0.034$. As a HBL, Markarian~501 possesses the synchrotron peak at X-ray energies, which makes it an ideal target for polarization studies in such energy band. However, the valley between the synchrotron and the VHE peak lies in the HE range at about a few MeV, so that Markarian~501 is not an ideal target for polarization analyses in the HE band. Concerning the emission mechanism, we consider only a leptonic scenario with typical HBL parameter values: $B^{\rm jet}_0 = 0.5 \, \rm G$, $y_{\rm em} = 3 \times 10^{16} \, \rm cm$ and $\gamma = 15$~\cite{tavMrk501}, since the hadronic emission mechanism cannot be applied owing to the high variability of the source. As the central $B^{\rm jet}_0$ is not so high, we cannot observe ALP-induced effects on photon polarization for an ALP mass $m_a = 10^{-10} \, \rm eV$ in the UV-X-ray band, as already pointed out in the previous two Subsections. This is the reason why we concentrate in the following only on the case $m_a \lesssim 10^{-14} \, \rm eV$. %We take an initial degree of linear polarization $\Pi_{L,0} = 0.3$ in all the UV-X-ray band according to~\cite{polar03}.
Since ALPs induce a modification to the final photon polarization, we choose an initial degree of linear polarization $\Pi_{L,0} = 0.3$ as an upper limit to $\Pi_{L,0}$~\cite{polar03}. %(see also note~\cite{noteMrk501}).

\begin{figure}
\centering
\includegraphics[width=0.5\textwidth]{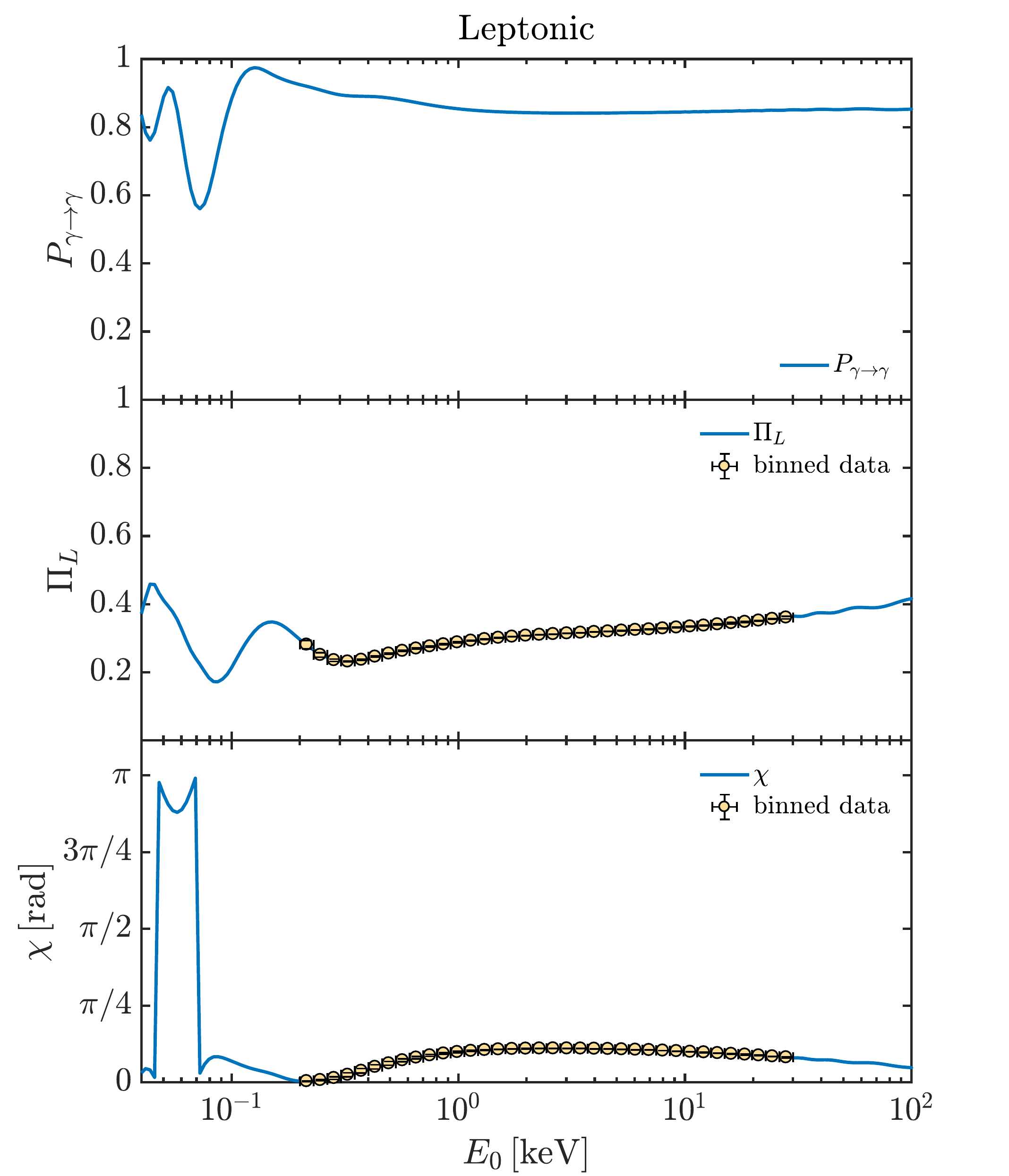}
\caption{\label{Mrk501AllPolKeV} Markarian~501: photon survival probability $P_{\gamma \to \gamma }$ (upper panel), corresponding final degree of linear polarization $\Pi_L$ (central panel) and final polarization angle $\chi$ (lower panel) in the energy range $(4 \times 10^{-2}-10^2) \, {\rm keV}$. We take $g_{a\gamma\gamma}=0.5 \times 10^{-11} \, \rm GeV^{-1}$, $m_a \lesssim 10^{-14} \, \rm eV$. The initial degree of linear polarization is $\Pi_{L,0}=0.3$.}
\end{figure}

\begin{figure}
\centering
\includegraphics[width=0.5\textwidth]{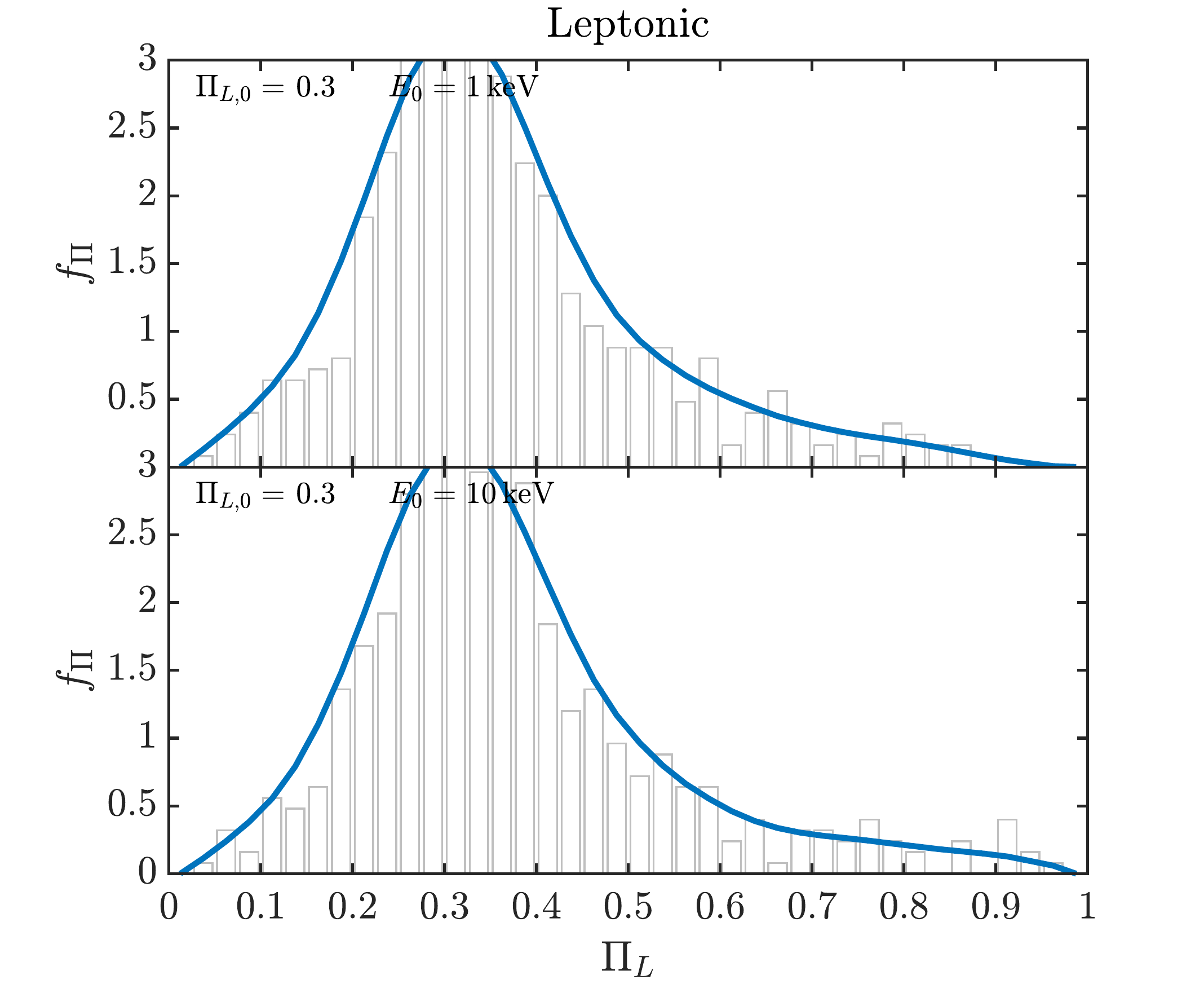}
\caption{\label{Mrk501DensKeV} Markarian~501: probability density function $f_{\Pi}$ arising from the plotted histogram for the final degree of linear polarization $\Pi_L$ at $1 \, \rm keV$ (upper panel) and $10 \, \rm keV$ (lower panel) by considering the system in Fig.~\ref{Mrk501AllPolKeV}. The initial degree of linear polarization is $\Pi_{L,0}=0.3$.}
\end{figure}

In Fig.~\ref{Mrk501AllPolKeV}, we report $P_{\gamma \to \gamma}$ and the corresponding final $\Pi_L$ and $\chi$. The photon-ALP system is in the weak mixing regime in the most of the considered energy band ($4 \times 10^{-2} \, {\rm keV}-10^2 \, \rm keV$) so that $P_{\gamma \to \gamma}$, $\Pi_L$ and $\chi$ turn out to be energy dependent. However, the final $\Pi_L$ appears not to be strongly modified by the photon-ALP interaction. In fact, the binned data in Fig.~\ref{Mrk501AllPolKeV} present a weak variability with little error bars. For both the two benchmark energies $E_0 = 1 \, \rm keV$ and $E_0 = 10 \, \rm keV$, Fig.~\ref{Mrk501DensKeV} confirms our previous statement: the behavior of $f_{\Pi}$ shows that the most probable value for the final $\Pi_L$ is still close to the initial one $\Pi_L \simeq \Pi_{L,0} = 0.3$ with a moderate broadening in the range $0.1 - 0.5$. %(see also note~\cite{noteMrk501}).
Note that Fig.~\ref{Mrk501DensKeV} is compatible with the recent IXPE results~\cite{IXPEmrk501}.

The reason for the different behavior of Markarian~501 with respect to the previously considered BL~Lacs is twofold: i) the lower value of its central $B^{\rm jet}_0$ with respect to the previous cases, which cannot substantially modify the photon polarization, ii) the relative proximity of Markarian~501 to the Earth, which implies a smaller broadening of the final photon polarization. For these reasons Markarian~501 does not appear as the best observational target for studies of ALP-induced effects on photon polarization in the X-ray band with observatories such as IXPE~\cite{ixpe}, eXTP~\cite{extp}, XL-Calibur~\cite{xcalibur}, NGXP~\cite{ngxp} and XPP~\cite{xpp}.

%\newpage

\subsection{1ES~0229+200}

1ES~0229+200 is the prototype of the so called extreme HBLs (EHBLs,~\cite{EHBL1,EHBL2}) and has been observed at redshift $z=0.1396$. Similarly to Markarian~501 it shows the synchrotron peak at X-ray energies, and it exhibits the valley between the synchrotron and the VHE peak in the HE range at about a few MeV. Therefore, 1ES~0229+200 represents an excellent observational target for polarization studies in the X-ray band but not in the HE range. We contemplate both the leptonic and the hadronic emission mechanisms. Correspondingly, we assume the typical EHBL parameter values $B^{\rm jet}_0 = 2 \, \rm mG$, $y_{\rm em} = 3 \times 10^{16} \, \rm cm$ and $\gamma = 50$ for the leptonic case and $B^{\rm jet}_0 = 0.5 \, \rm G$, $y_{\rm em} = 3 \times 10^{16} \, \rm cm$ and $\gamma = 15$ regarding the hadronic one~\cite{EHBL1,EHBL2}. In the following, we investigate only the case $m_a \lesssim 10^{-14} \, \rm eV$ for the same reasons explained for Markarian~501: the central $B^{\rm jet}_0$ is not high enough to allow -- in the UV-X-ray band -- sizable ALP-induced effects on photon polarization for an ALP mass $m_a = 10^{-10} \, \rm eV$. We assume an initial degree of linear polarization $\Pi_{L,0} = 0.3$ in the whole UV-X-ray band and for both the leptonic and the hadronic scenarios according to~\cite{polar03}.

\begin{figure*}
\centering
\includegraphics[width=0.867\textwidth]{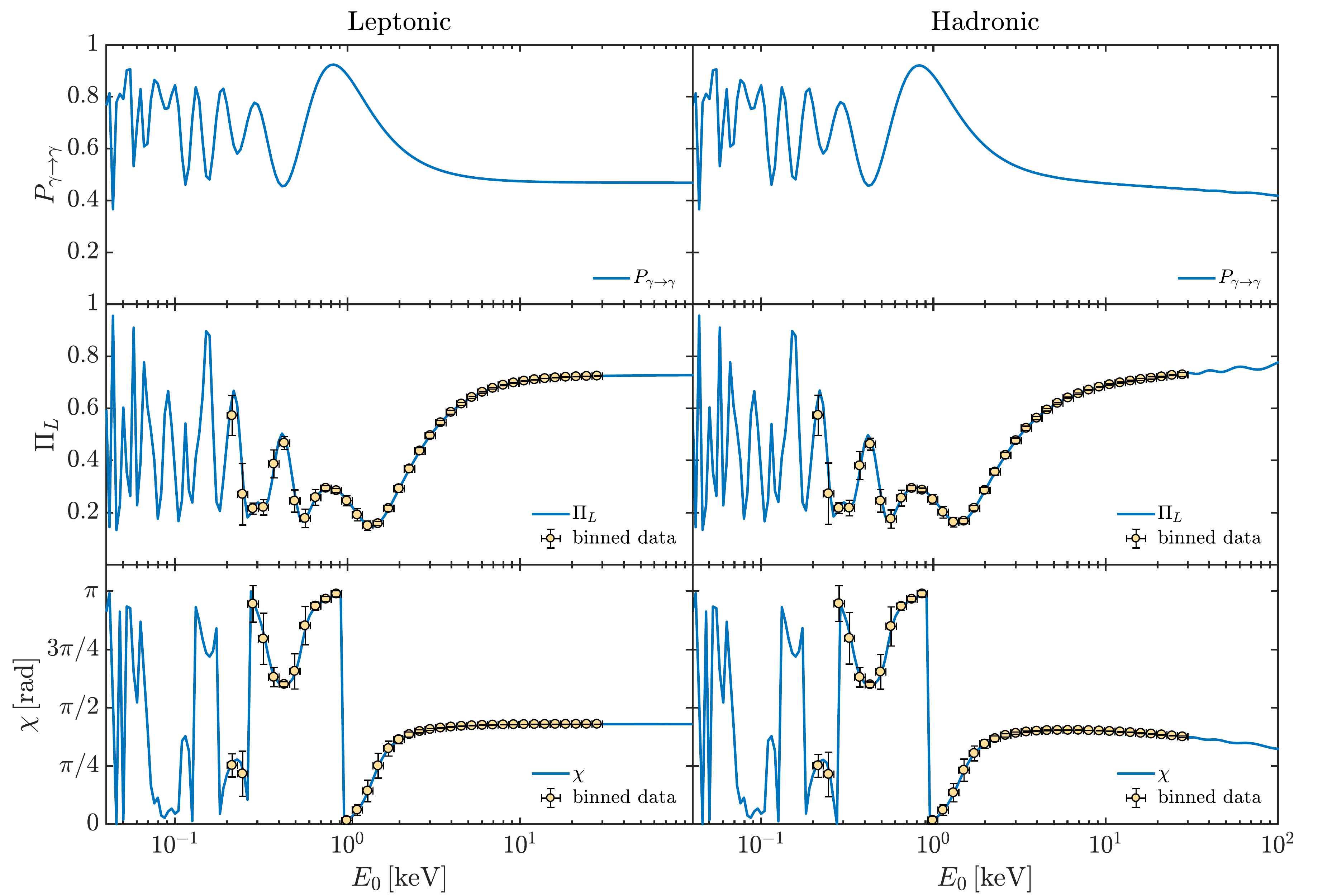}
\caption{\label{1ES0229AllPolKeV} 1ES~0229+200: photon survival probability $P_{\gamma \to \gamma }$ (upper panels), corresponding final degree of linear polarization $\Pi_L$ (central panels) and final polarization angle $\chi$ (lower panels) in the energy range $(4 \times 10^{-2}-10^2) \, {\rm keV}$. We take $g_{a\gamma\gamma}=0.5 \times 10^{-11} \, \rm GeV^{-1}$, $m_a \lesssim 10^{-14} \, \rm eV$. We consider a leptonic and hadronic emission mechanism in the left and right column, respectively. The initial degree of linear polarization is $\Pi_{L,0}=0.3$ both in the leptonic and hadronic cases.}
\end{figure*}

\begin{figure*}
\centering
\includegraphics[width=0.867\textwidth]{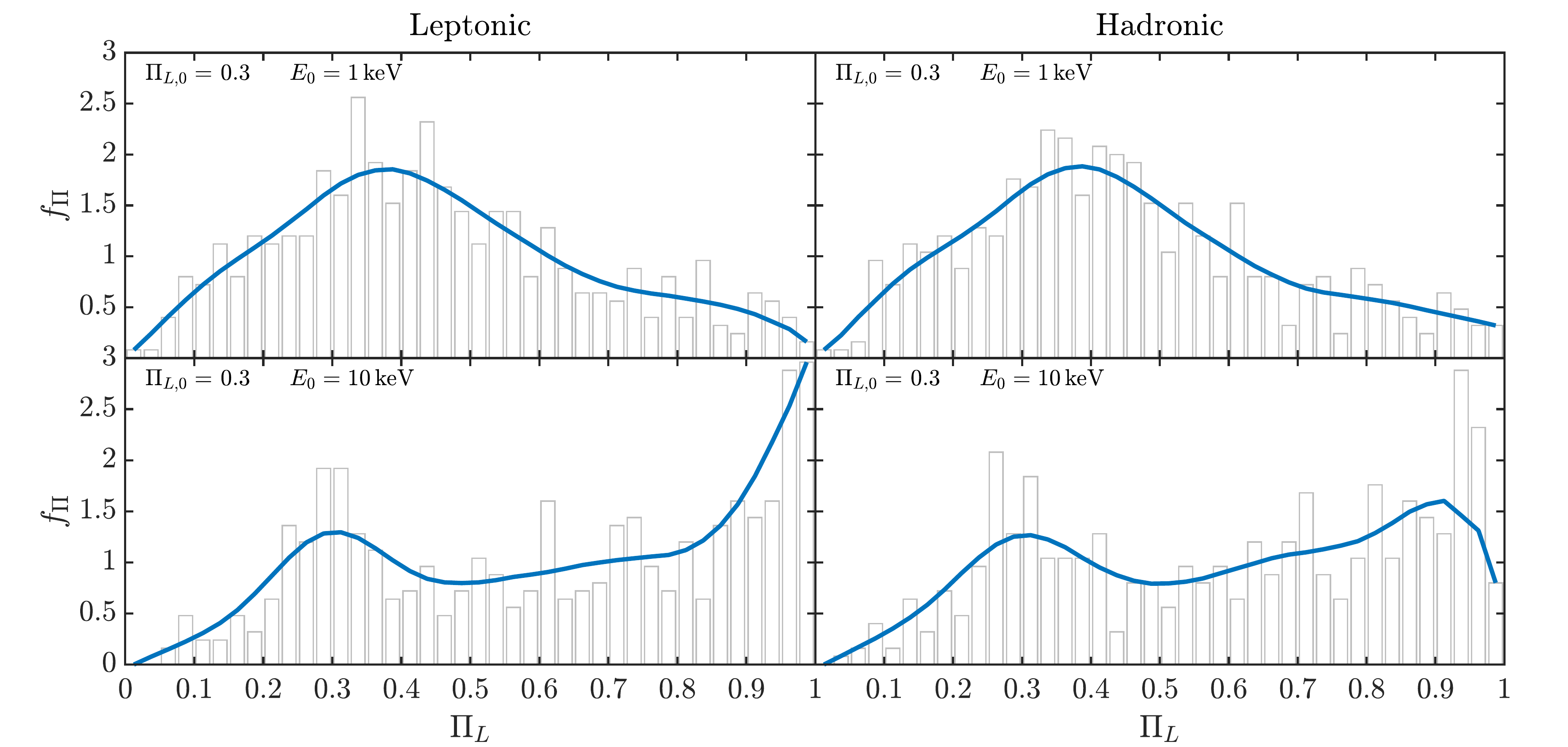}
\caption{\label{1ES0229DensKeV} 1ES~0229+200: probability density function $f_{\Pi}$ arising from the plotted histogram for the final degree of linear polarization $\Pi_L$ at $1 \, \rm keV$ (upper panels) and $10 \, \rm keV$ (lower panels) by considering the system in Fig.~\ref{1ES0229AllPolKeV}. We consider a leptonic and hadronic emission mechanism in the left and right column, respectively. The initial degree of linear polarization is $\Pi_{L,0}=0.3$ both in the leptonic and hadronic cases.}
\end{figure*}

In Fig.~\ref{1ES0229AllPolKeV}, we report $P_{\gamma \to \gamma}$, and the corresponding final $\Pi_L$ and $\chi$ for both the leptonic and the hadronic cases. What is evident is that the two scenarios are very similar and we observe a little variation only for $E_0 \gtrsim 5 \, \rm keV$. The reason for this behavior is as follows. Since the value of the central $B^{\rm jet}_0$ is very low as compared to all other considered cases, we see that the ALP-induced effects are negligible inside the jet. We can observe ALP-induced effects in the jet for $E_0 \gtrsim 5 \, \rm keV$ only in the hadronic case because of the higher value of the central $B^{\rm jet}_0$. ALP-induced effects on photon polarization is therefore dominated by photon-ALP conversion in the extragalactic space and in the Milky Way. This fact explains why the leptonic and hadronic scenarios are so similar. In the UV-X-ray band the photon-ALP beam propagates in the weak mixing regime, so that all quantities in Fig.~\ref{1ES0229AllPolKeV} and, in particular, the final $\Pi_L$ are energy dependent. The binned data in Fig.~\ref{1ES0229AllPolKeV} show low error bars since the pseudo-oscillatory behavior is moderately variable.

The behavior of $f_{\Pi}$ in Fig.~\ref{1ES0229DensKeV} confirms that the leptonic and hadronic scenarios are very similar. For the benchmark energy $E_0 = 1 \, \rm keV$ we find a large broadening of the initial $\Pi_{L,0} = 0.3$ but the most probable final $\Pi_L$ remains $\Pi_L = 0.3$. Instead, for $E_0 = 10 \, \rm keV$ we observe that the most probable final $\Pi_L$ is $\Pi_L > 0.8$ with large broadening.

Because conventional physics cannot predict a final $\Pi_L > 0.8$, 1ES~0229+200 appears as a favored observational target for studies concerning ALP-induced effects on photon polarization for observatories such as IXPE~\cite{ixpe}, eXTP~\cite{extp}, XL-Calibur~\cite{xcalibur}, NGXP~\cite{ngxp} and XPP~\cite{xpp} especially for $E_0 \gtrsim 5 \, \rm keV$.

\subsection{Discussion about polarization detectability}

As discussed above, OJ~287 and BL~Lacertae appear as good observational targets for studies about ALP effects on photon polarization both in the X-ray and in the HE band, since we find that the most probable value for the final $\Pi_L$ turns out to be $\Pi_L \gtrsim 0.8$ in some cases, and this value cannot be attained even in the case of the hadronic emission mechanism, as shown in Fig.~\ref{Polin}. Because of the low flux in the HE band, 1ES~0229+200 is a good candidate for ALP studies in the X-ray range only. Instead, Markarian~501 -- which cannot be considered for the HE range for the same reason of 1ES~0229+200 -- turns out not to be a good observational target even in the X-ray band because of its proximity to the Earth and of its low central jet magnetic field.

We have also checked the impact of the photon-ALP conversion in the extragalactic space: while $B_{\rm ext}=1 \, \rm nG$ appears as the most probable value (see Sec. III.D), we have explored also the case $B_{\rm ext} < 10^{-15} \, \rm G$, which leads to a negligible photon-ALP interaction. 

In particular, when $m_a \lesssim 10^{-14} \, {\rm eV}$, the photon-ALP conversion in the extragalactic space is effective for $B_{\rm ext} =1 \, \rm nG$: a $B_{\rm ext} < 10^{-15} \, \rm G$ -- with corresponding inefficient photon-ALP interaction -- produces a decrease in the width of the broadening of $f_{\Pi}$ in the previous figures. Such an effect is lower in the case of Markarian~501, since it is very close to us. We only note some sizable modification in the behavior of $f_{\Pi}$ for 1ES~0229+200 at $10 \, \rm keV$. In particular, high values of $f_{\Pi}$ are not anymore the most probable ones when $B_{\rm ext} < 10^{-15} \, \rm G$. The reason for that lies in the fact that the photon-ALP conversion in the jet of 1ES~0229+200 is not as strong as in any other considered BL~Lac for the low value of its central $B^{\rm jet}_0$.

Instead, since photon-ALP conversion in the extragalactic space is not very efficient even for $B_{\rm ext}=1 \, \rm nG$ when $m_a = 10^{-10} \, {\rm eV}$ especially in the UV-X-ray band, our results remain substantial unchanged by assuming $B_{\rm ext}< 10^{-15} \, \rm G$.

The contribution of the photon-ALP conversion in the jet depends on the behavior of ${\bf B}^{\rm jet}_0$: in the case of a high $B^{\rm jet}_0$, the photon-ALP conversion in the jet produces a substantial effect, while its contribution decreases with a lower $B^{\rm jet}_0$. The effect of the photon-ALP interaction in other crossed regions is less important.    

In our previous analysis, we have considered two cases about the ALP parameter space ($m_a, g_{a\gamma\gamma}$): (i) $[m_a \lesssim 10^{-14} \, {\rm eV} , g_{a\gamma\gamma} = 0.5 \times 10^{-11} \, {\rm GeV^{-1}}]$; (ii) $[m_a = 10^{-10} \, {\rm eV}, g_{a\gamma\gamma} = 0.5 \times 10^{-11} \, {\rm GeV}^{-1}]$. Both these choices are within the CAST bound~\cite{cast}. However, some new studies about photon-ALP conversion inside galaxy clusters suggest that the case $[m_a \lesssim 10^{-14} \, {\rm eV} , g_{a\gamma\gamma} = 0.5 \times 10^{-11} \, {\rm GeV^{-1}}]$ is disfavored with respect the case $[m_a = 10^{-10} \, {\rm eV}, g_{a\gamma\gamma} = 0.5 \times 10^{-11} \, {\rm GeV}^{-1}]$, as shown in~\cite{limFabian,limJulia,limKripp,limRey2}.

In a previous paper concerning ALP effects on photon polarization in galaxy clusters~\cite{grtClu}, we have concluded that the HE range represents the best choice for such studies, as the case disfavored by the latter bounds ($[m_a \lesssim 10^{-14} \, {\rm eV} , g_{a\gamma\gamma} = 0.5 \times 10^{-11} \, {\rm GeV^{-1}}]$) is the only one producing effects in the X-ray band. However, this conclusion does not apply to BL~Lacs for two reasons: (i) we observe ALP effects in the X-ray band also in the case $[m_a = 10^{-10} \, {\rm eV}, g_{a\gamma\gamma} = 0.5 \times 10^{-11} \, {\rm GeV}^{-1}]$ in the hadronic emission mechanism for OJ~287 and BL~Lacertae; (ii) the bounds~\cite{limFabian,limJulia,limKripp,limRey2} are derived by considering galaxy clusters for particular choices of the cluster magnetic field ${\bf B}^{\rm clu}$ morphology. If the ${\bf B}^{\rm clu}$ configuration is such that the photon-ALP conversion is extremely inefficient, we could observe no effects due to the unfavorable ${\bf B}^{\rm clu}$ morphology, instead of concluding by rejecting the case $[m_a \lesssim 10^{-14} \, {\rm eV} , g_{a\gamma\gamma} = 0.5 \times 10^{-11} \, {\rm GeV^{-1}}]$. Therefore, since in the present situation we are dealing with sources different from those considered in~\cite{limFabian,limJulia,limKripp,limRey2}, we can neither exclude nor consider as disfavored even the case $[m_a \lesssim 10^{-14} \, {\rm eV} , g_{a\gamma\gamma} = 0.5 \times 10^{-11} \, {\rm GeV^{-1}}]$.

As a result, we can expect ALP-induced polarization effects both in the X-ray band and in the HE range, which can be detectable by observatories such as IXPE~\cite{ixpe}, eXTP~\cite{extp}, XL-Calibur~\cite{xcalibur}, NGXP~\cite{ngxp}, XPP~\cite{xpp} and COSI~\cite{cosi}, e-ASTROGAM~\cite{eastrogam1,eastrogam2}, AMEGO~\cite{amego}, respectively. In addition, we want to stress that many other possibilities about the choice of the ALP parameter space ($m_a, g_{a\gamma\gamma}$) can be explored, but they produce final results similar to those arising from our benchmark values.

\section{Conclusions}

In this paper, we have investigated the ALP-induced effects on the final degree of linear polarization $\Pi_L$ and on the polarization angle $\chi$ and we have analyzed the probability density function $f_{\Pi}$ of $\Pi_L$ associated with several realizations of the photon-ALP beam propagation process, for photons emitted at the jet base of a few  BL~Lacs: OJ~287, BL~Lacertae, Markarian~501 and 1ES~0229+200. We have used the state-of-the-art knowledge about the astrophysical parameters of the media (blazar jet, host galaxy, extragalactic space, Milky Way) crossed by the photon-ALP beam and realistic values of the initial degree of linear polarization $\Pi_{L,0}$. In addition, we have considered both the leptonic and the hadronic emission mechanisms. By specializing the procedure developed in~\cite{galantiPol} to generic blazars, we have analyzed two scenarios concerning the ALP parameter space ($m_a, g_{a\gamma\gamma}$): (i) $[m_a \lesssim 10^{-14} \, {\rm eV} , g_{a\gamma\gamma} = 0.5 \times 10^{-11} \, {\rm GeV^{-1}}]$; (ii) $[m_a = 10^{-10} \, {\rm eV}, g_{a\gamma\gamma} = 0.5 \times 10^{-11} \, {\rm GeV}^{-1}]$. These choices are within the CAST bound~\cite{cast}. We have obtained ALP-induced features on the final $\Pi_L$, which can be detected by IXPE~\cite{ixpe}, eXTP~\cite{extp}, XL-Calibur~\cite{xcalibur}, NGXP~\cite{ngxp} and XPP~\cite{xpp} in the X-ray band, and by COSI~\cite{cosi}, e-ASTROGAM~\cite{eastrogam1,eastrogam2} and AMEGO~\cite{amego} in the HE range. Our findings can be summarized as follows.

\begin{enumerate}[(i)]

\item In the X-ray band, the major effects are produced in the case $[m_a \lesssim 10^{-14} \, {\rm eV} , g_{a\gamma\gamma} = 0.5 \times 10^{-11} \, {\rm GeV^{-1}}]$. In particular, OJ~287 and BL~Lacertae show a broadening of the initial $\Pi_{L,0}$, while 1ES~0229+200 represents a very good observational target since its most probable value for the final $\Pi_L$ is $\Pi_L > 0.8$ at $E_0 = 10 \, \rm keV$. The latter value cannot be explained within conventional physics, and so a detection would represent an additional hint at the ALP existence. We have no sizable difference between leptonic and hadronic emission models for 1ES~0229+200. In the case $[m_a = 10^{-10} \, {\rm eV}, g_{a\gamma\gamma} = 0.5 \times 10^{-11} \, {\rm GeV}^{-1}]$, the photon-ALP conversion is inefficient and negligible in all crossed media apart from the blazar jet, provided that hadronic models with a high central magnetic field $B^{\rm jet}_0$ are considered. Accordingly, we observe a dimming of the initial $\Pi_{L,0}$ concerning OJ~287 and BL~Lacertae. In all other situations, the photon-ALP interaction is so inefficient that we observe neither effects on nor modifications of the initial $\Pi_{L,0}$.

\item In the HE range, we find strong signals of ALP-induced effects on $\Pi_L$ and $\chi$ in both cases $[m_a \lesssim 10^{-14} \, {\rm eV} , g_{a\gamma\gamma} = 0.5 \times 10^{-11} \, {\rm GeV^{-1}}]$ and $[m_a = 10^{-10} \, {\rm eV}, g_{a\gamma\gamma} = 0.5 \times 10^{-11} \, {\rm GeV}^{-1}]$, and for both the leptonic and the hadronic emission mechanisms. Here, we propose OJ~287 and BL~Lacertae as good observational targets for ALP studies concerning polarization. In particular, the most probable value for the final $\Pi_L$ is $\Pi_L \gtrsim 0.8$ especially around $3 \, \rm MeV$. As discussed above, the detection of such a high value of the final $\Pi_L$ would be a strong indication in favor of the ALP existence, as conventional physics cannot explain it.

\end{enumerate}

Instead, Markarian~501 turns out not to be a good observational target for studies about ALP-induced polarization effects, as explained in Sec. IV.C.

Both the two considered cases about the ALP parameter space are promising and the case $[m_a \lesssim 10^{-14} \, {\rm eV} , g_{a\gamma\gamma} = 0.5 \times 10^{-11} \, {\rm GeV^{-1}}]$ cannot a priori be excluded, as explained in Sec. IV.E.

Lorentz invariance violation (LIV) can produce a variation in the final $\Pi_L$ but in terms of a decrease of the initial $\Pi_{L,0}$ only~\cite{LIVpol}. Thus, in all cases where the polarization gets increased with respect to conventional physics, the explanation can only be due to the photon-ALP interaction.

Complementary tests on ALPs similar to those considered in this paper can be performed by the new generation of VHE gamma-ray observatories like CTA~\cite{cta}, HAWC~\cite{hawc}, GAMMA-400~\cite{g400}, LHAASO~\cite{lhaaso}, TAIGA-HiSCORE~\cite{desy} and HERD~\cite{herd}. Moreover, laboratory experiments like the upgrade of ALPS II at DESY~\cite{alps2}, the planned IAXO~\cite{iaxo,iaxo2} and STAX~\cite{stax}, the techniques developed by Avignone and collaborators~\cite{avignone1,avignone2,avignone3} and the planned ABRACADABRA experiment~\cite{abracadabra} -- in case ALPs were the bulk of the dark matter -- can detect them.

\section*{Acknowledgments}

We thank Enrico Costa for discussions. G.G. acknowledges a contribution from the grant ASI-INAF 2015-023-R.1. M.R. acknowledges the financial support by the TAsP grant of INFN. F.T. acknowledges a contribution from the grant INAF Main Stream project ‘High-energy extragalactic astrophysics: towards the Cherenkov Telescope Array’. This work was made possible also by the funding by the INAF Mini Grant `High-energy astrophysics and axion-like particles', PI: Giorgio Galanti.


\begin{thebibliography}{20}
\bibitem{footnote1} Our goal is to provide an idea of what an observer may expect, so that the central values of the data points are calculated by averaging the sampled calculation points within the bin and the uncertainties are evaluated by calculating the standard deviation. See also~\cite{grtClu} for the complete description of the assumed binning procedure.% While a fully correct procedure implies to build a Monte Carlo code to calculate events based on an emitted spectrum, we do not expect sizable differences since the spectrum is not so steep to provide different weights at the different energies.
\end{thebibliography}
\end{document}